\documentclass[letterpaper]{article}
\usepackage{aaai}
\usepackage{times}
\usepackage{helvet}
\usepackage{courier}
\usepackage{color}
\usepackage{url}
\usepackage{verbatim} 
\usepackage{subfigure} 
\usepackage{multirow}
\usepackage{xspace}
\usepackage{graphicx}
\usepackage{epsfig}
\usepackage{amsmath}
\usepackage{amssymb}
\usepackage{bm}

\newcommand{\denselistA}{ \itemsep -0.5pt\topsep-5pt\partopsep-7pt }

\newcommand{\hide}[1]{}
\newcommand{\xhdr}[1]{\vspace{1.7mm}\noindent{{\bf #1.}}}

\newcommand{\eg}{\emph{e.g.}}
\newcommand{\ie}{\emph{i.e.}}

\begin{document}

\title{Quantifying Information Overload in Social Media and\\ its Impact on Social Contagions}

\author{
Manuel Gomez-Rodriguez\\
MPI for Intelligent Systems \\
\tt{manuelgr@tue.mpg.de} \\
\And 
Krishna P. Gummadi\\
MPI for Software Systems \\
\tt{gummadi@mpi-sws.org} \\
\And
Bernhard Sch\"{o}lkopf\\
MPI for Intelligent Systems \\
\tt{bs@tue.mpg.de} \\
}

\maketitle

\begin{abstract}
\begin{quote}
Information overload has become an ubiquitous pro\-blem in modern
society. Social media users and microbloggers receive an endless flow
of information, often\- at a rate far higher than their cognitive
abilities to process the information.
In this paper, we conduct a large scale quantitative study of
information overload and evaluate its impact on information
dissemination in the Twitter social media site.
We model social media users as information processing systems that
queue incoming information according to some policies, process information 
from the queue at some {\it unknown} rates and decide to forward some of 
the incoming information to other users.
We show how timestamped data about tweets received and forwarded by
users can be used to uncover key properties of their queueing policies and
estimate their information processing rates and limits. Such an understanding
of users'{} information processing behaviors allows us to infer
whether and to what extent users suffer from information overload.

Our analysis provides empirical evidence of information processing
limits for social media users and the prevalence of information
overloading.
The most active and popular social media users are often the ones that
are overloaded.
Moreover, we find that the rate at which users receive information
impacts their processing behavior, including how they prioritize
information from different sources, how much information they process,
and how quickly they process information.
Finally, the susceptibility of a social media user to social
contagions depends crucially on the rate at which she receives
information. An exposure to a piece of information, be it an idea, a
convention or a product, is much less effective for users that receive
information at higher rates, meaning they need more exposures to adopt
a particular contagion.
\end{quote}
\end{abstract}

\section{Introduction} \label{sec:introduction}
\label{sec:intro}
\noindent Since Alvin Toffler popularized the term ``Information overload" in his bestselling 1970 book Future Shock~\cite{toffler1984future}, it has become a major problem in modern society. 
Information overload occurs when the amount of input to a system
exceeds its processing capacity. Humans have li\-mi\-ted cognitive
processing capacities, and consequently, when they are overloaded with
information, their quality of decision making
suffers~\cite{gross1964managing}.

The advent of social media and online social networking has led to a
dramatic increase in the amount of information a user is exposed to,
greatly increasing the chances of the user experiencing an information
overload. In particular, microbloggers complain of information
overload to the greatest\- extent. Surveys show that two thirds of
Twi\-tter users have felt that they receive too many posts, and over
half of Twi\-tter users have felt the need for a tool to filter out the
irrelevant posts~\cite{bontcheva2013}. Prior stu\-dies show that
information overload has a major impact on users in a broad range of
domains: work productivity~\cite{mckinsey2011},
re\-commen\-dation systems~\cite{borchers1998}, or information
systems~\cite{bawden2009}. However, all these prior works have
typically relied on qualitative analysis, surveys or small-scale
experiments.

In this paper, we perform a large-scale quantitative study of
information overload experienced by users in the Twitter social media
site. The key insight that enables our study is that users'{} information
processing behaviors can be reverse engineered through a careful
analysis of the times when they receive a piece of information and
when they choose to forward it to other users.
Abstractly, we think of a Twitter user as an information processing
system that queues the in\-co\-ming flow of information (or in-flow) 
according to a last-in first-out (LIFO) policy, processes the information
from the queue at some \emph{unknown} rate, and decides to forward
some of the processed information.
By uncovering key properties of the queueing policies and processing rates of
users with diffe\-rent in-flow rates, we attempt to (i) understand how
users'{} information processing behaviors vary with the in-flow rate,
(ii) estimate information processing limits of users, and (iii)
ultimately infer the level of information overload on users.

To this end, we use data gathered from Twitter, which comprises of all
public tweets published during a three months period, from July 2009
to September 2009.
In Twi\-tter, a user'{}s information queue co\-rres\-ponds to her
Twitter timeline (or feed), the incoming flow of information co\-rres\-ponds to the
stream of tweets published by the users she follows, and the queuing
policy corresponds to the way the user reads and forwards (retweets)
tweets from her feed.
We show how we can reverse engineer different characteristics of the
users'{} processing behaviors using only the timestamps of the tweets
sent / received and the social graph between the users.

Our analysis yields several insights that not only reveal the extent
to which users in social media are overloaded with information, but
also help us in understanding how information overload influences
users'{} decisions to forward and disseminate information to other
users: 
\begin{enumerate}
\denselistA
\item We find empirical evidence of a limit on the amount of
information a user produces per day. We observed that very few 
Twitter users produce more than $\sim$40 tweets/day.

\item In contrast to the tight limits on information produced, we
find no limits on the information received by Twitter users. The
information received scales linearly with the number of users followed
and many Twitter users follow several hundreds to thousands of other
users.

\item We observed a threshold rate of incoming information ($\sim$30
tweets/hour), below which the probability that a user forwards any
received tweet holds nearly constant, but above which the probability
that a user forwards any received tweet begins to drop
substantially. We argue that the threshold rate roughly approximates
the limit on information processing capacity of users and it allows us
to identify users that are overloaded.

\item When users are not overloaded, we observe that they tend to
process and forward information faster as the rate at which they
receive information increases. However, when users are overloaded, we
find that the higher the rate at which they receive information, the
longer the time they take to process and forward a piece of
information.

\item When users are overloaded, they appear to prioritize tweets from a
selected subset of sources. We observe that even as overloaded users
follow more users, the set of users whose tweets they forward remains
fairly constant and increases very slowly.

\end {enumerate}

Our findings about the information processing and forwarding behaviors
of overloaded users have important implications for propagation of
information via {\it social contagions}. While a large number of
previous studies have focused on social contagions in on-line
media~\cite{du13nips,goel2012structure,manuel10netinf,leskovec2009kdd}, most ignore the impact of
information overload on users'{} forwarding behavior. They consider
the spread of a single piece of information through an un\-der\-lying
network, ignoring the effect of lots of other background
information simultaneously spreading through the network. Here, we
investigate the impact of such background traffic on the spread of
social contagions. In particular, we study the propagation of
hashtags~\cite{romero11twitter}, retweet
conventions~\cite{kooti2012emergence}, and URL shortening service
usage~\cite{antoniades2011we} through the Twitter network. One of our
main findings is that the rate at which users receive information, i.e.,
the extent to which they are overloaded, has a strong impact on the
number of exposures a user needs to adopt a contagion.

\section{Related Work}
\noindent The work most closely related ours~\cite{backstrom2011center,hodas2012visibility,miritello2013limited} 
investigates the impact that the amount of social ties of a social media user has on the way she interacts or exchanges 
information with her friends, followees or contacts.
Backstrom et al. measure the way in which an individual divides his or her attention across 
contacts by analyzing Facebook data. Their analysis suggests that some people focus most of their attention on a small 
circle of close friends, while others disperse their attention more broadly over a large set.
Hodas et al. quantify how a user'{}s limited attention is divided among information sources (or 
followees) by tracking URLs as mar\-kers of information in Twitter. They provide empirical evi\-dence that highly connected 
individuals are less likely to propagate an arbitrary tweet.
Miritello et al. analyze mobile phone call data and note that individuals exhibit a finite commu\-ni\-cation
capacity, which limits the number of ties they can maintain active.
The common theme is to investigate whether there is a limit on the amount of ties (\eg, friends, followees or phone contacts) 
people can maintain, and how people distribute attention across them.
\begin{figure*}[t]
	\centering
	\subfigure[Out-flow rate]{\includegraphics[width=0.24\textwidth]{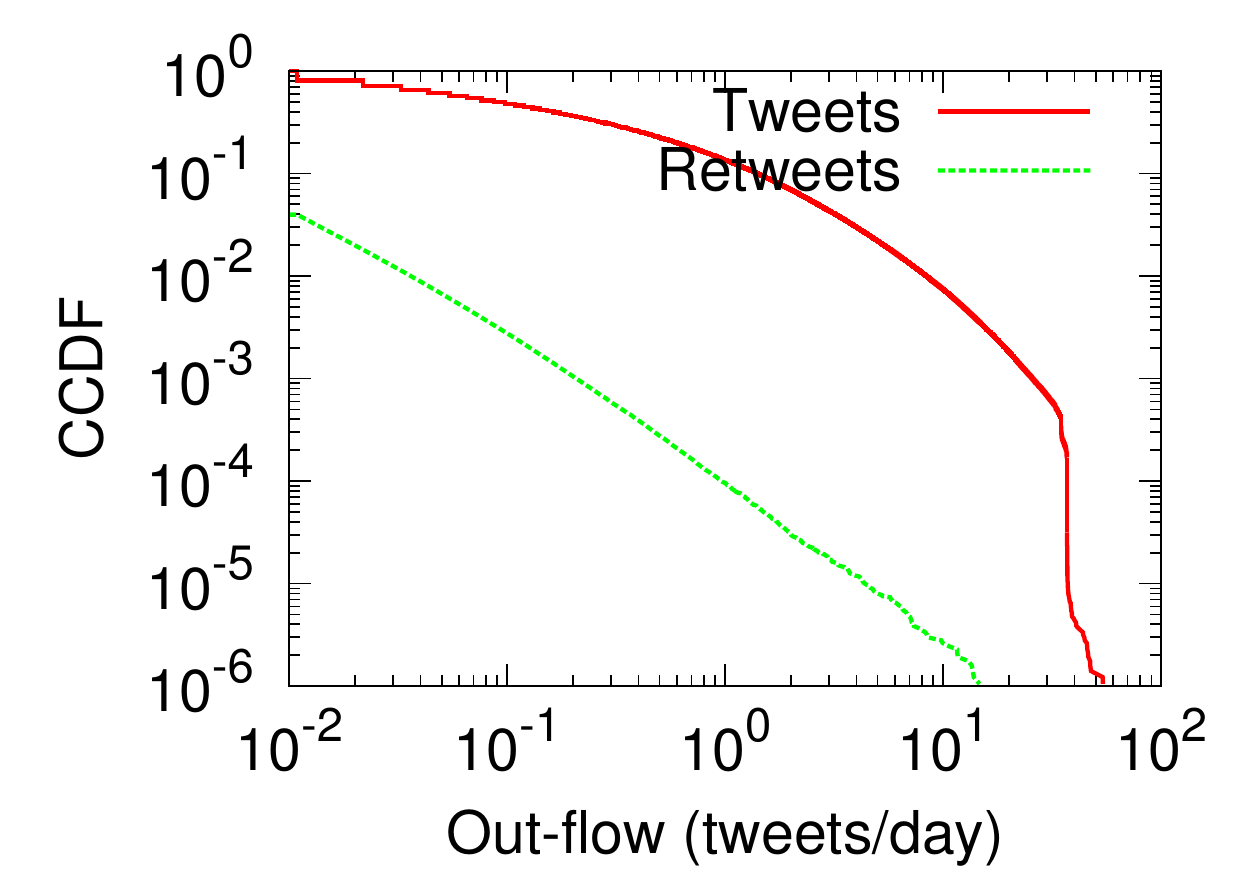}\label{fig:out-flow-rate-ccdf}}
	\subfigure[In-flow rate]{\includegraphics[width=0.24\textwidth]{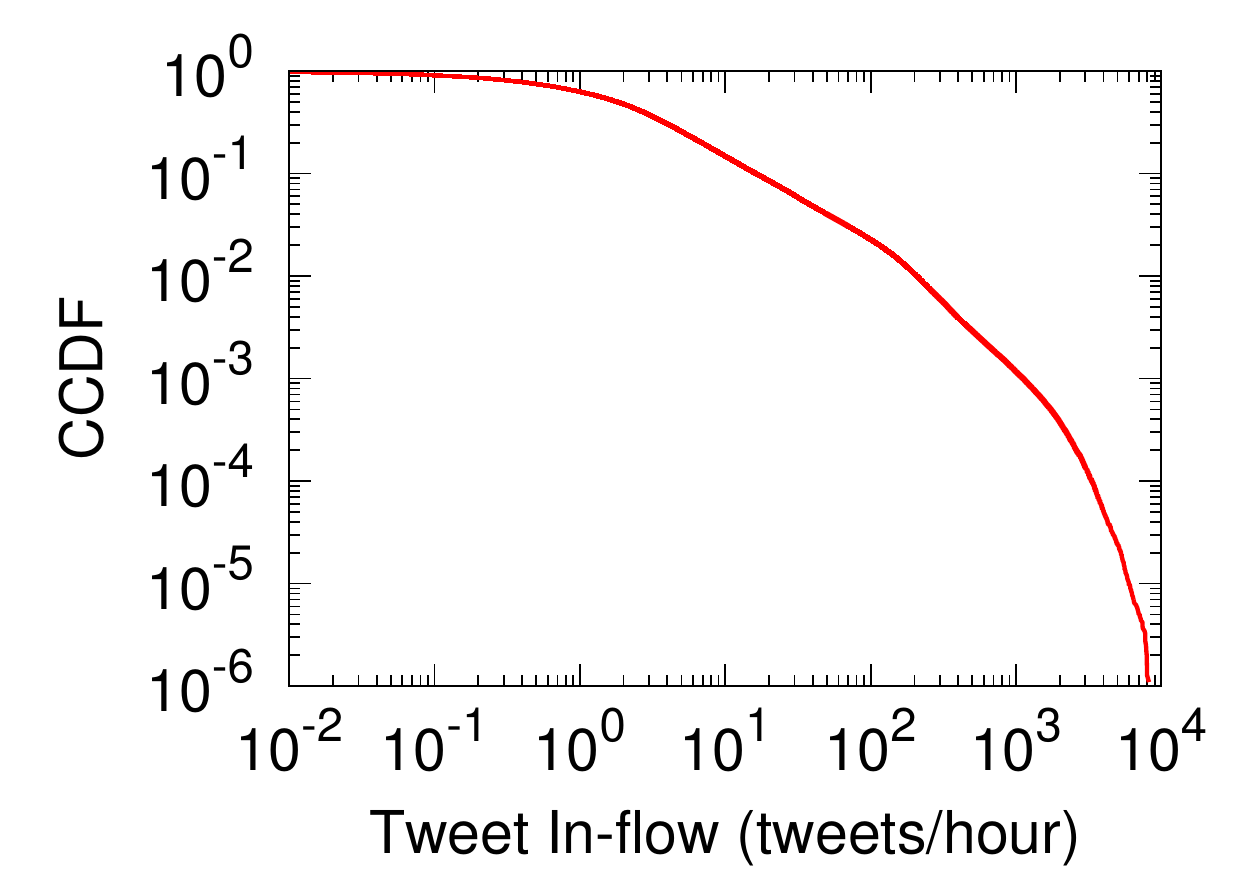}\label{fig:in-flow-rate-ccdf}}
	\subfigure[Followees]{\includegraphics[width=0.24\textwidth]{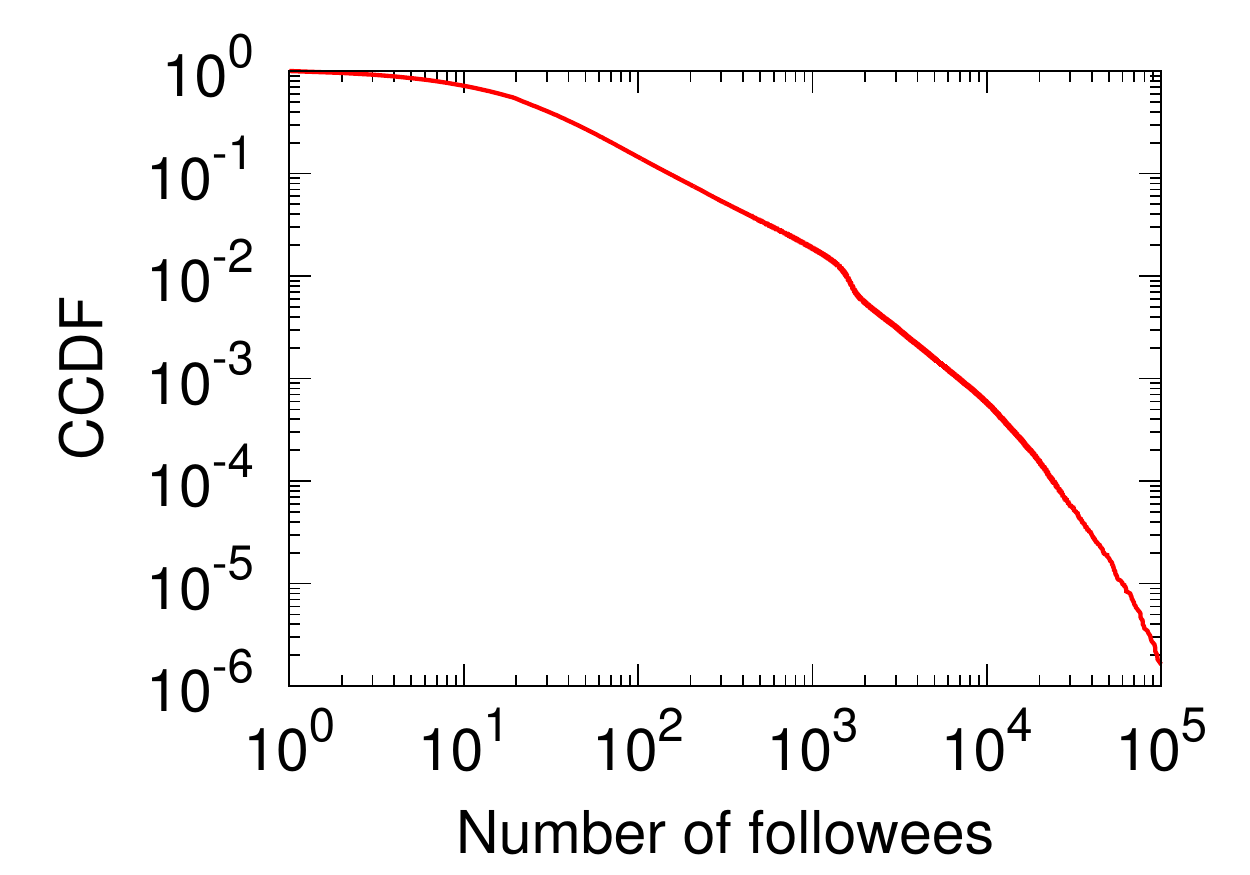}\label{fig:followees-ccdf}}
	\subfigure[In-flow rate vs. followees]{\includegraphics[width=0.24\textwidth]{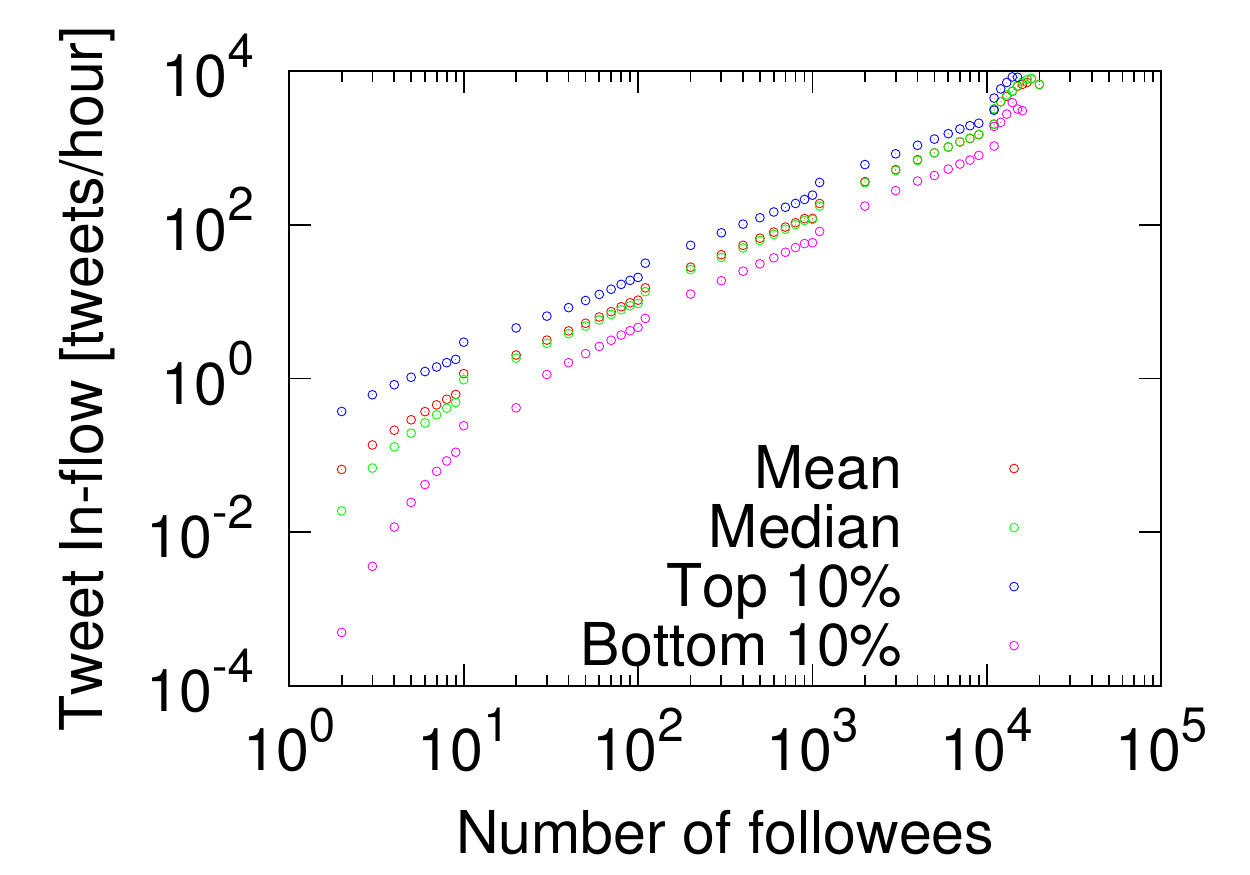}\label{fig:in-flow-rate-number-followees}}
	\caption{Distributions of tweet out-flow rates, tweet in-flow rates and followees.} \label{fig:basic-distributions}
\end{figure*}

However, partly due to a lack of complete temporal data, there are several fundamental differences between the above mentioned studies and our 
work.
First, we find a sharp thres\-hold or phase transition that roughly approximates the limit on the amount of information social media users can process (read and possibly 
forward) (Fig.~\ref{fig:in-flow-probability-retweet-b}). This allows us to identify users that are overloaded.
Instead, no sharp threshold exists on the number of ties a user can distribute attention across~\cite[Fig. 3]{hodas2012visibility}. 
Second, we uncover key properties of the users'{} reading and for\-war\-ding policies and find dramatic qualitative and quantitative differences between non overloaded
and overloaded users. These differences have important implications for the large and growing number of studies on information propagation and social 
contagion.
In contrast, previous studies could not unveil most of these differences. 
Third, our work explores to which extent the probability of adoption of an idea, social convention, or product, or more generally, a contagion, depends on the social 
media user'{}s in-flow and finds stri\-king differences on the susceptibility for social contagion of users with different in-flows.
Instead, Backstrom et al. and Miritello et al. do not investigate social contagion; Hodas et al. only pays attention to one type of contagion, urls, and analyzes to 
which extent the probability of forwarding a url depends on the number of ties a user distribution attention across.
Finally, our work naturally extends existing models of information and influence propagation to support background traffic. The extended models are able to capture 
two important features for information cascades: cascade sizes are limited and a few cascades have prolonged lifetimes.
In contrast, previous work does not provide any roadmap or ideas on how to incorporate their findings in terms of number of ties to well-known information and influence 
propagation models.

Last, very recently, there have been attempts to analyze and model information propagation assuming competition and cooperation between contagions~\cite{goyal2012competitive,myers12clash,weng2012competition}. However, this line of work considers only interactions between pairs of contagions~\cite{myers12clash} 
or present theoretical models without experimental validation~\cite{goyal2012competitive}.
In contrast, our work investigates the effect\- that observing a great number of contagions simultaneously propagating, not necessarily informative or in\-te\-res\-ting, has on the 
ability of a person to process and forward information.
We believe this is a key point for understanding information propagation in the age of information overload.

\section{Methodology} \label{sec:methodology}

\noindent In this section, we describe our methodology for es\-ti\-mating information processing limits and behaviors of users in social media
sites from observational data.
One of the key cha\-llen\-ges we faced is that we could not directly observe (i.e., gather data about) how users read, process and forward
information.
Instead, we only had access to observational data about the times when users receive each piece of information and the times when they
forward a particular piece of information. However, we do not know which pieces of information each user actually reads or when the users
read the pieces of information.
To overcome this limitation, our methodology assumes each user applies some policy to read, process and forward information.

We think abstractly of a social media user as an information processing system that receives an incoming flow of information (or
in-flow) on an information queue.
The in-flow is composed of information generated by other information processing systems (users), which the system decides to 
subs\-cribe to.
Then, we assume the system applies a last-in first-out (LIFO) queueing policy to process and sometimes forward particular pieces of
information from the in-flow.
The rationale behind our LIFO queuing policy choice is twofold. First, in most social media platforms, each user has a feed (be it in the
form a Facebook user'{}s wall, a Twi\-tter user'{}s timeline or an Instagram user'{}s feed) which she periodically checks out, where the
information generated by other users she subscribes to is accumulated.
Second, a user browses the feed from top to bottom because, even if the user skips some of the content, she still has to scroll from the
top to the bottom of the feed. Therefore, it seems natural to consider the feed as a LIFO queue.
In Twitter, the information queue is persistent and co\-rres\-ponds to a user'{}s Twi\-tter timeline, the incoming flow of information (or
in-flow) corresponds to the stream of tweets published by the user'{}s followees\footnote{Followees of a Twitter user are the users she
follows on Twi\-tter.}, and the queuing policy corresponds to the way a user reads and possibly retweets tweets from her feed.
Importantly, at the time when we gathered our data (in the year 2009), Twitter sorted each user'{}s feed in inverse chronological order and
thus, it is possible to reconstruct a user'{}s queue at any given time.

In this framework, we use the times in which information was generated and forwarded to estimate properties of the queueing such as the
queues sizes (i.e., the position in the queue beyond which users do not process information), the queues processing delays (i.e., the time
spent by information in the queue waiting to be processed), or the prio\-ri\-ty given to information received from each followee on the users'{} 
queues, among others.
It may seem surprising that we are able to estimate such properties without observing additional behavioral data such as which information
each user actually reads or when the user reads each piece of information.
Our key insight is that retweets (forwarded tweets) provide a sample of the underlying queueing policies.
More specifically, we assume that a user retweets soon (almost ins\-tan\-ta\-neous\-ly) after reading a tweet for the first time. Under this
assumption, we observe that (i) retweet delays are independent samples of the reading delays, and, (ii) queue positions of retweeted 
tweets at the time of retweets are lower bounds of the queues sizes. 
In subsequent sections, we show how the above observations make it possible to estimate information processing limits and behaviors of
users from our limited observational data consisting only of the times when users receive and forward tweets.

\subsection{Twitter Dataset}

\noindent We use data gathered from Twitter as reported in previous work~\cite{cha2010measuring}, which comprises the following 
three types of information: profiles of $52$ million users, $1.9$ bi\-llion directed follow links among these users, and $1.7$ bi\-llion public tweets posted 
by the collected users. The fo\-llow link information is based on a snapshot taken at the time of data collection, in September 2009.
In our work, we limit ourselves to tweets published from July 2009 to September 2009 and filter out users that did not tweet before June 2009, in order 
to be able to consider the social graph to be \emph{approximately} static. 
After the preprocessing steps, we have 5,704,427 active users, 563,880,341 directed edges, 318,341,537 tweets and 1,807,748 retweets.

For every user, we build her in-flow and out-flow by co\-llec\-ting all tweets published by the people she follows, the tweets she published and the retweets she 
generated.
Since back in 2009, Twitter did not have a retweet button, and ins\-tead, users were explicitly identifying retweets using se\-ve\-ral conventions, we identify
retweets by searching the most common conventions, following previous work~\cite{kooti2012emergence}.
Another important characteristic of Twitter in 2009 is that it did not have features such us ``Lists'' and ``Per\-so\-na\-lized Suggestions'' and so the primary way 
users received and processed information was through their feed, for which we have complete data. However, this comes at the cost of observing a smaller 
number of users and information flows.

\section{Information Processing Limits}
\label{sec:processing-limit}
\noindent In this section, we investigate the information processing capacity of social media users. More specifically, we attempt to answer the following three questions in the context 
of the Twitter social medium: 

\begin{enumerate}
\denselistA
\item What, if any, are the limits on the amount of information users generate and forward to their friends? 
\item Do users try to limit the amount of information they receive from other users?
\item Is there evidence of users suffering from information overload? If yes, does information overload impact users decisions to forward and disseminate information? 
\end{enumerate}

\begin{figure}[t]
	\centering
	\subfigure[Retweet rate vs in-flow rate]{\includegraphics[width=0.23\textwidth]{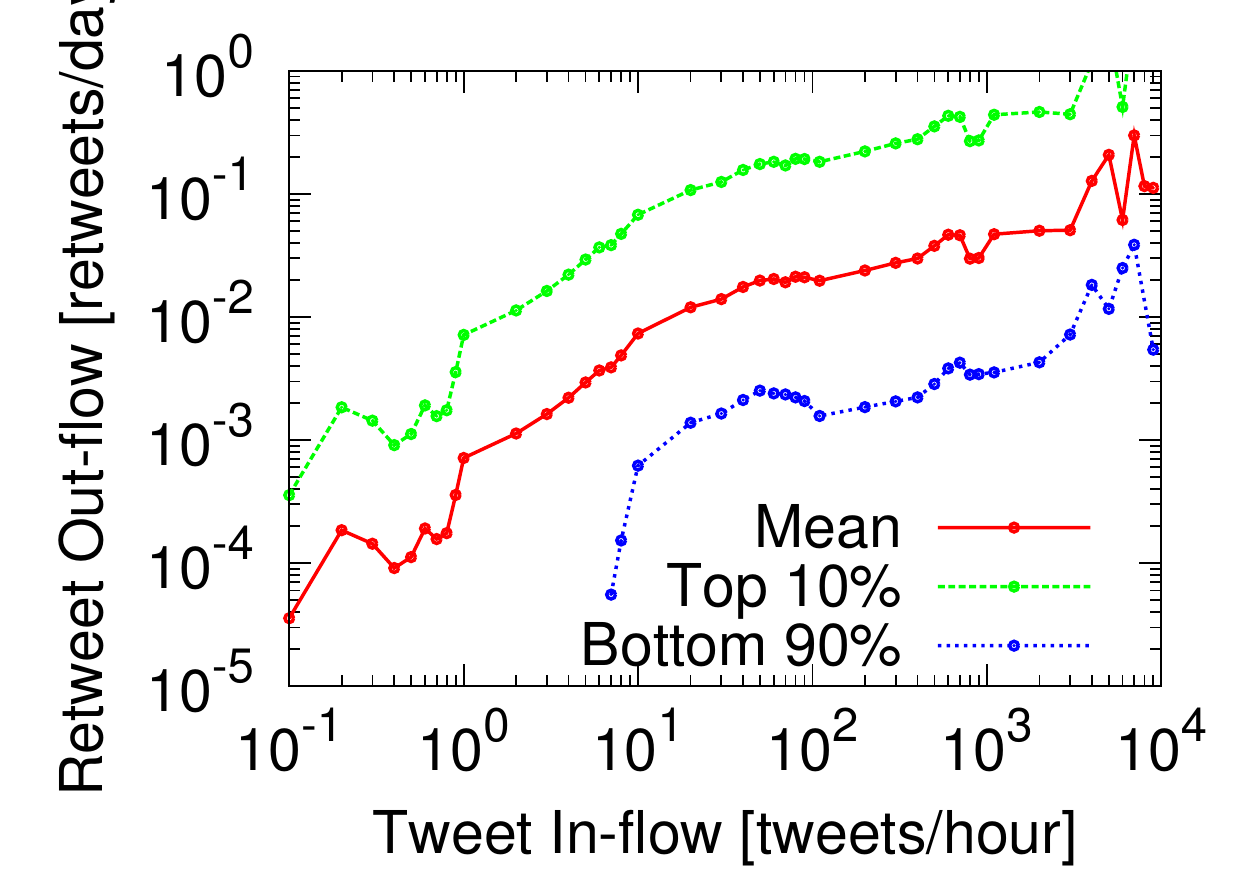}}
	\subfigure[Retweet prob. vs in-flow rate]{\includegraphics[width=0.23\textwidth]{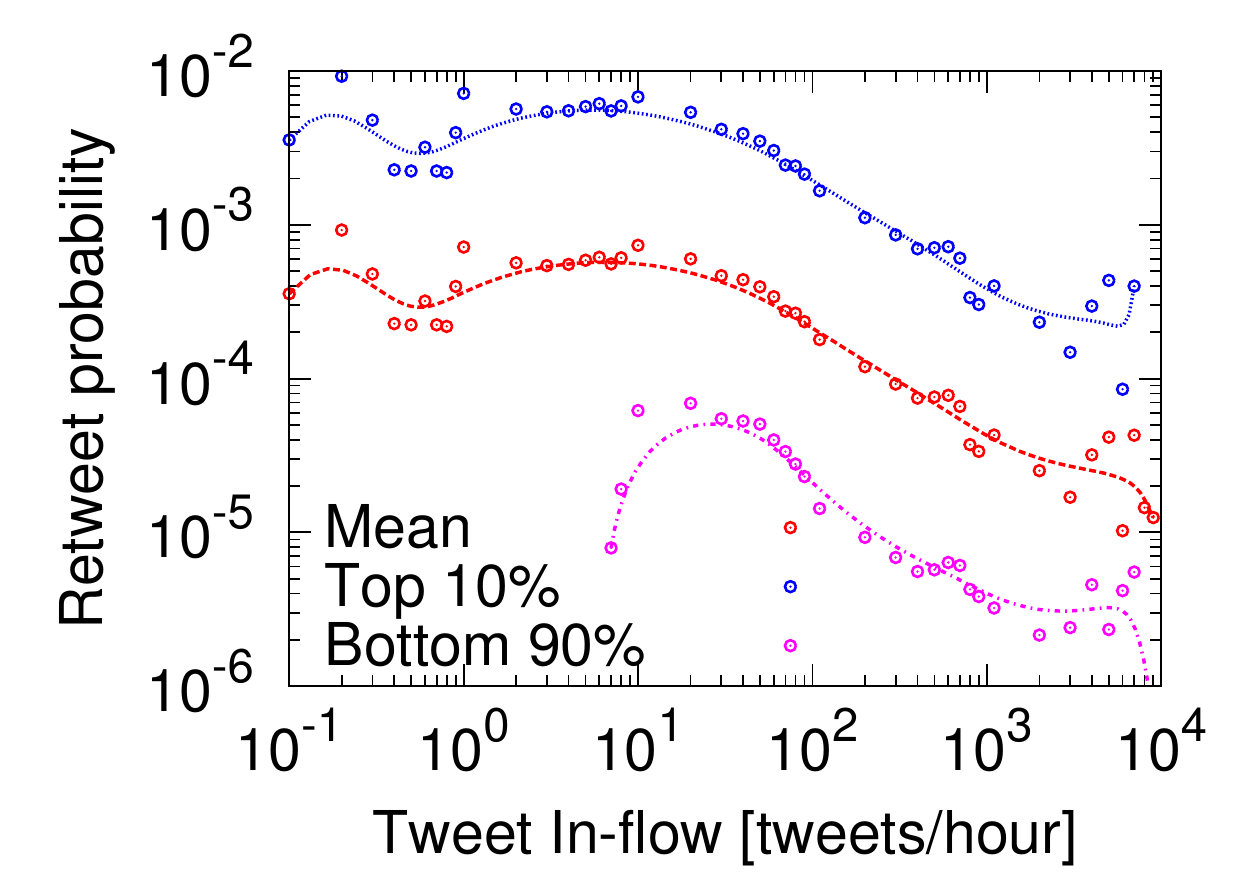}\label{fig:in-flow-probability-retweet-b}}
 	\caption{Retweets vs in-flow. Panel (a) shows retweet rate and panel (b) shows probability of retweet.} \label{fig:in-flow-probability-retweet}
\end{figure}

\subsection{Limits on information generation \& forwarding}

\noindent We begin by focusing on the rate at which users produce information in Twitter. Specifically, we compute the number 
of tweets they generate per day, considering original tweets and retweets of the tweets they receive from other users se\-pa\-ra\-tely.  
Figure~\ref{fig:out-flow-rate-ccdf} shows the distributions of tweet {\it out-flow rates}, i.e., the number of total tweets and retweets 
in\-di\-vi\-dual Twitter users post per day. We find several interesting patterns in the plots which we briefly discuss next.

First, we observe a sharp fall-off on the total number of tweets a user produces per day, located at $\sim$40 tweets/day. 
This provides empirical evidence that there exists a limit on the amount of tweets, or more generally, information, most Twitter users 
produce every day. 
There are several plausible reasons that may explain the existence of such limit. For example, posting tweets is a manual task and therefore, 
intuitively, one would expect to find some limit on the number of tweets posted by a user. Moreover, users may choose to limit their posts for 
fear of overloading (or spamming) their followers with information~\cite{comarela2012understanding}.

Second, the distribution of the number of retweets a user produces per
day follows a power-law, in contrast with the distribution of the
total number of tweets a user produces per day, which decays much
slower up to a sharp fall-off. This suggests that the underlying
policies users follow to process and forward incoming tweets, or more
generally, information, are essentially different to the policies they
follow to publish their own tweets.

\subsection{Origins of information overload}
\noindent We now shift our attention to the amount of information Twitter users receive per time unit. Figure~\ref{fig:in-flow-rate-ccdf} shows the
distribution of tweet {\it in-flow rates}, i.e., the number of tweets individual Twitter users receive per hour. While 50\% of users receive
fewer than 50 tweets per day, 10\% of users receive more than 500 tweets per day. Many of these tweets contain URLs (pointers to web
pages)~\cite{mislove2007measurement}. Surprisingly, unlike in the case of tweet out-flows, users do not appear to be limiting their tweet in-flows
(potentially, by following fewer users). The relatively high in-flow rates of tweets and information for some of the users suggests the
potential for users to be overwhelmed with information.

Next we investigate whether users receiving more (less) information are systematically following higher (smaller) number of
users. Figure~\ref{fig:in-flow-rate-number-followees} shows the average, median, and top / bottom 10\% in-flow rates against number of
followees for Twitter users. The plot shows that the tweet in-flow rates of individual users are strongly and linearly correlated with
the number of users they follow. As one might expect, the higher the number of followees a user follows the greater the number of tweets
she receives. Figure~\ref{fig:followees-ccdf} shows the distribution of number of followees for Twitter users. Similar to what has been
observed in prior work~\cite{kwak2010twitter}, we find that roughly 30\% of the users follow more than 50 people, but only around 
0.15\% of the users follow more than 5,000 people.

Our findings suggest that the origins of information overload in social media lie in the tendency of Twitter users to oversubscribe,
i.e., follow a lot more users than those whose tweets they can process (as we show in the next section). {\it Why do users choose to receive
a lot more information than they can process?} One explanation proposed by prior work is that users are choosing to follow users but
not ne\-ce\-ssa\-ri\-ly all their tweets, i.e., users' following behavior is driven by their desire to follow people they find interesting, but
they are not necessarily interested in all the tweets posted by the users they follow~\cite{hodas13icwsm}. Thus, information overload in social media
today arises out of users' tendency to socialize (exchange information) with many other users.

\subsection{Evidence of information overload} \label{sec:overload}
\noindent Our above observations about unbounded information in-flow rates and bounded information 
out-flow rates for users suggest that the chances of a user forwarding an incoming piece of information may depend on
the in-flow rate itself, i.e., the extent to which the user is overloaded. Here, we elucidate this question by investigating 
the relationship between the tweet in-flow rate and the retweet out-flow rate.

Figure~\ref{fig:in-flow-probability-retweet}(a) shows how the retweet out-flow rate $\lambda_r$ varies against the tweet in-flow rate
$\lambda$. Interestingly, as the in-flow rate increases, the retweeting rate increases, but at a de\-clining rate. In other words,
the retweet rate seems to follow a law of diminishing returns with respect to the in-flow rate; mathematically,
$\lambda_r(\lambda_1+\Delta\lambda)
- \lambda_r(\lambda_1) \leq \lambda_r(\lambda_2+\Delta\lambda)
- \lambda_r(\lambda_2)$ for $\lambda_1 \geq \lambda_2$.
Figure~\ref{fig:in-flow-probability-retweet}(b) shows how the probability of retweeting an incoming tweet $\beta_r$ varies against
the in-flow rate $\lambda$. Interestingly, we find two different in-flow rate regimes: below $\sim$30 tweets/hour, the retweeting probability 
$\beta_r$ is relatively constant, however, over $\sim$30 tweets/hour, it falls sharply against the inflow rate, specifically, we observe a 
power law $\beta_r \propto \lambda^{-0.65}$, where we found the power-law coefficient using maximum likelihood estimation (MLE). 
The second regime represents the scenario in which a user is overloaded. In this scenario, the more tweets a user receives, the lower 
the probability of any received tweet to be retweeted by the user, indicating a greater information overload on the user. 
Note that this phenomena cannot be simply explained by the existence of a limit on the users'{} out-flow rates -- even if the retweet probability would 
remain cons\-tant, the retweet rate would still be far away from the limit on information generation we have observed previously. 
For example, consider there exists a user with an incoming rate $\lambda = 10^3$ tweets/hour and a retweet probability $\beta_r = 10^{-3}$, then her 
retweet rate would be $1 \ll 40$ retweets/hour.
Therefore, we argue it provides empirical evidence of information overloading.
Perhaps surprisingly 10\% of the active users in our dataset have an in-flow rate in the second regime and thus are likely suffering from 
information overload.
Who are those overloaded users? They are typically very popular users, with a significant amount of followers, who tweet more frequently 
than the \emph{average} user. For example, while 95\% of the overloaded users have more than $200$ followers, 86\% of the non overloaded 
users have less than $100$ fo\-llo\-wers. Strikingly, the set of overloaded users are responsible for 45\% of all retweets.
Our results provide strong empirical evidence for (a) the existence of an information processing 
limit for social media users and (b) the prevalence of information overloading, i.e., breaching of
the information processing limits, in social media today.

Our above finding has important implications for the large and growing number of studies on information 
cascades and social contagion in social media~\cite{du13nips,goel2012structure,manuel10netinf,leskovec2009kdd}. 
In particular, many studies today focus on the di\-sse\-mi\-nation of a single piece of information, 
completely ig\-no\-ring the presence of background traffic. 
On the contrary, our finding here suggests that excessive background traffic can have a strong negative impact on 
information dissemination and it provides supporting empirical evidence for the few prior studies that have 
postulated that information over\-loading might explain why most information cascades in social media fail to reach 
epidemic proportions~\cite{hodas2012visibility,ver2011stops}.
Later, we will investigate further how information cascades and social contagion depend on the background traffic.
\begin{figure}[t]
	\centering
	\subfigure[Queue position distribution]{\includegraphics[width=0.23\textwidth]{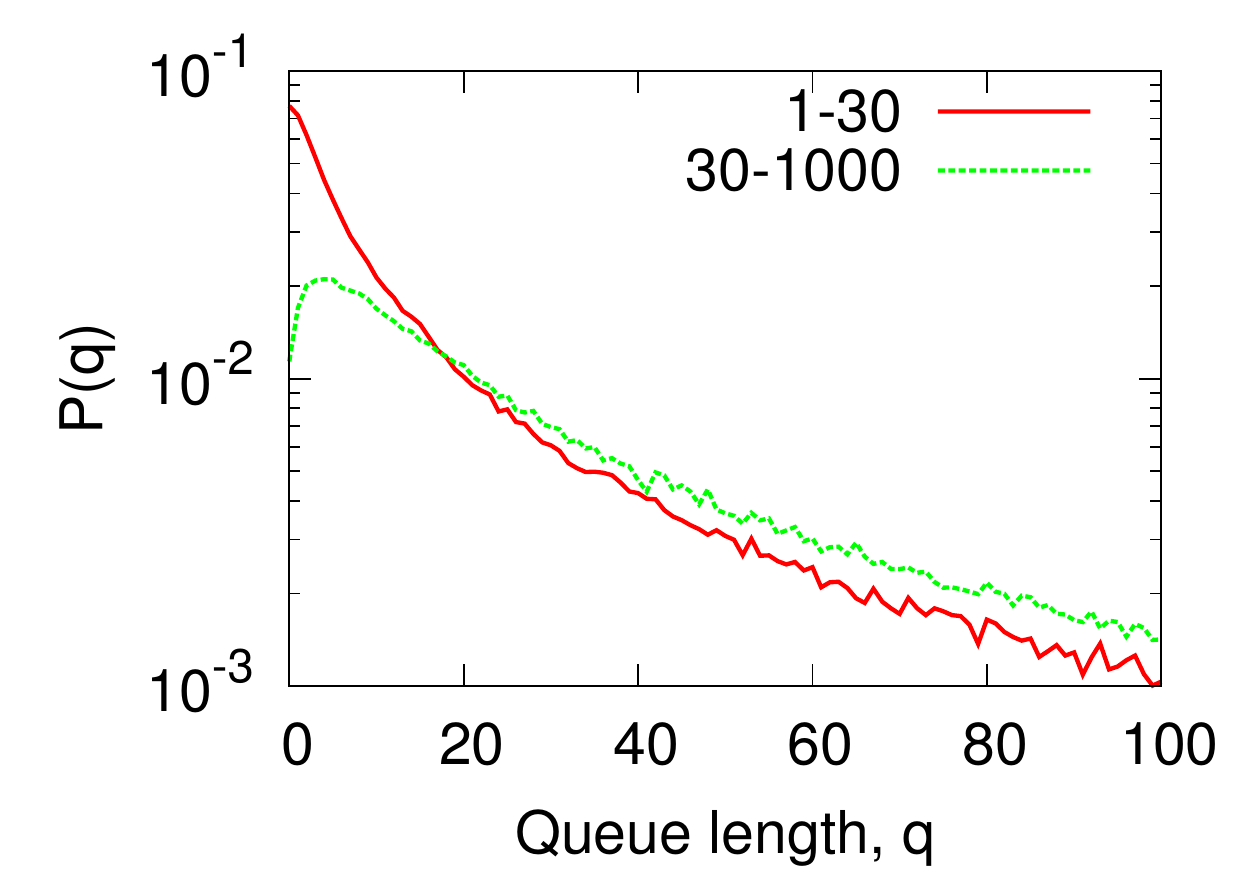}}
	\subfigure[Average queue position]{\includegraphics[width=0.23\textwidth]{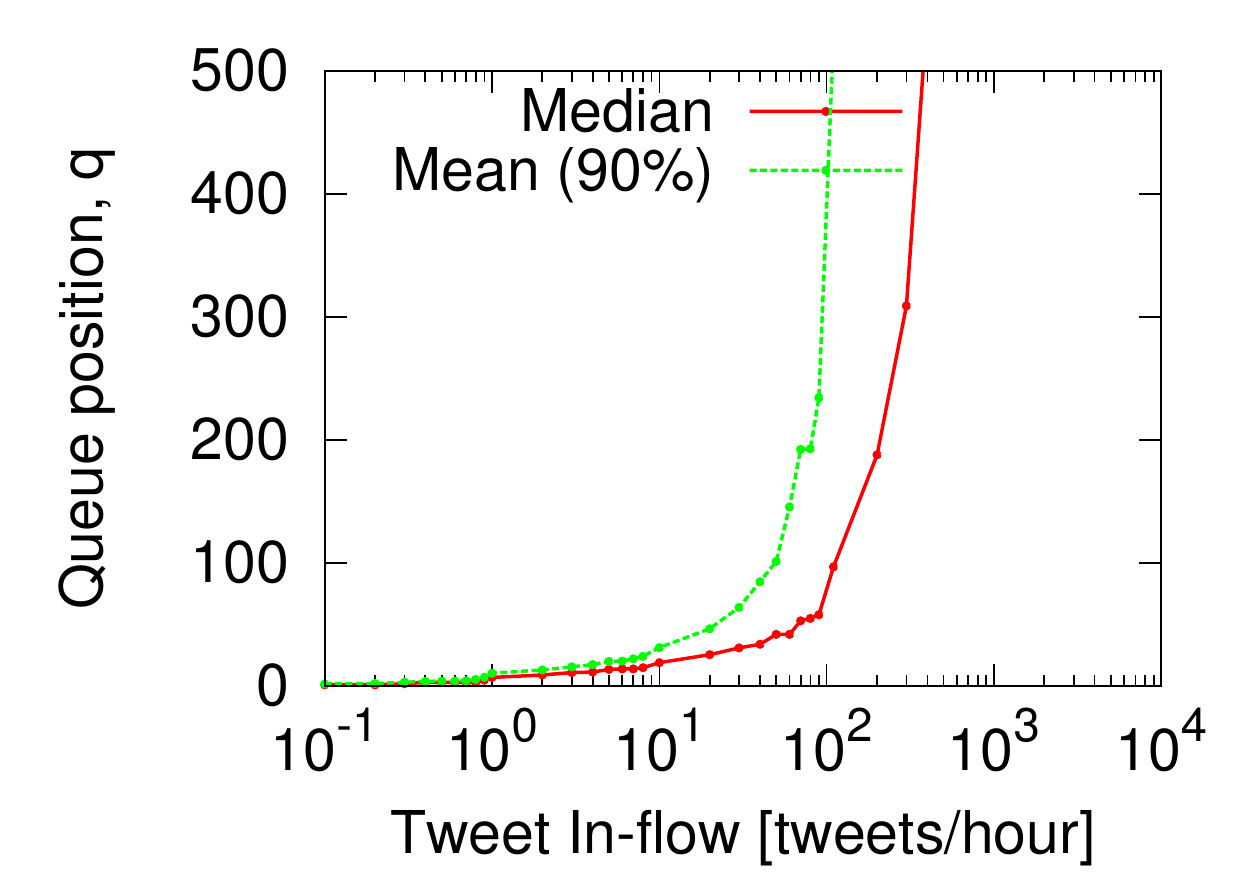}}
	\caption{Queue position. Panel (a) shows the empirical distribution and panel (b) shows average/median position of a tweet on the user'{}s queue (feed) at the 
	time when it got retweeted for different in-flow rates, where $q = 0$ means the tweet was at the top of the user'{}s queue at retweeting time.} \label{fig:queue position}
\end{figure}

\section{User Processing Behaviors}
\label{sec:information-queues}
\noindent In the previous section, we showed evidence of information overload amongst Twitter users, but lacked a detailed understanding
of the underlying ways in which users process the information they receive. As discussed earlier, information received by a user can be thought 
of as being added to her LIFO information queue and processed asynchronously whenever the user logs into the system. In this section, we 
present a detailed analysis of how users process information in their queues. Specifically, we attempt to characterize three aspects of their 
information processing behaviors:

\begin{enumerate}
\denselistA
\item How does the probability of processing a piece of information vary with its position in the queue? Can we estimate the queue sizes
  beyond which users effectively ignore information?
\item How large are typical queueing (processing) delays for information? How does it vary with the rate of incoming
  information?
\item When users are overloaded, do they tend to separate information into priority queues? Do they tend to 
  prioritize and process information from certain users over others?
\end{enumerate}

\subsection{Queue position vs. processing probability}
\noindent Intuitively, we may expect that when Twitter users login and begin processing tweets in their queues (feeds), they are much more likely to process 
tweets closer to the head of the queue than the tweets further down. Here, we estimate the likelihood that users process information further down in the queue 
by estimating the position of the tweets that are retweeted at the time of their retweet.
\begin{figure*}[t]
	\centering
	\subfigure[Delay (forwarded information)]{\includegraphics[width=0.28\textwidth]{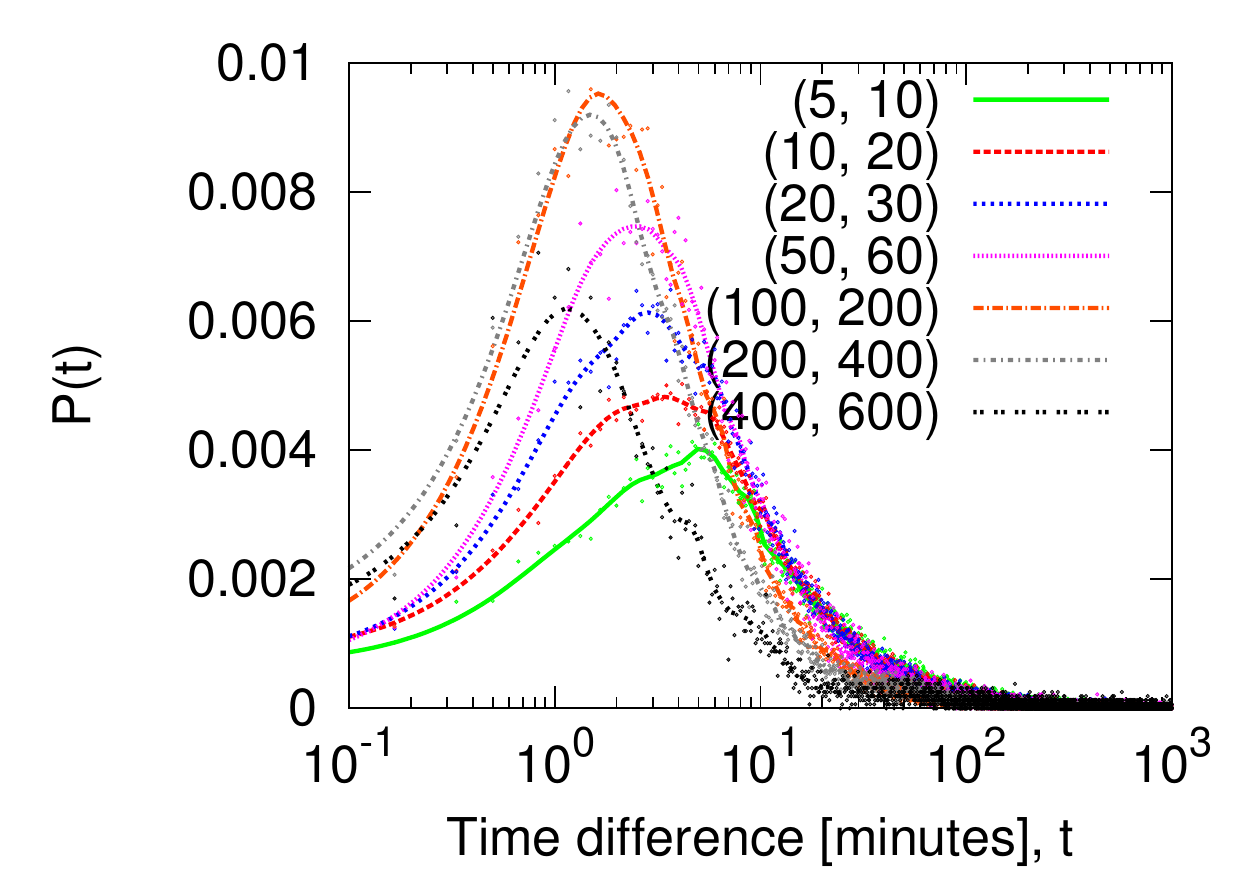}} \hspace{2mm}
	\subfigure[Delay vs in-flow (forwarded information)]{\makebox[6cm][c]{\includegraphics[width=0.28\textwidth]{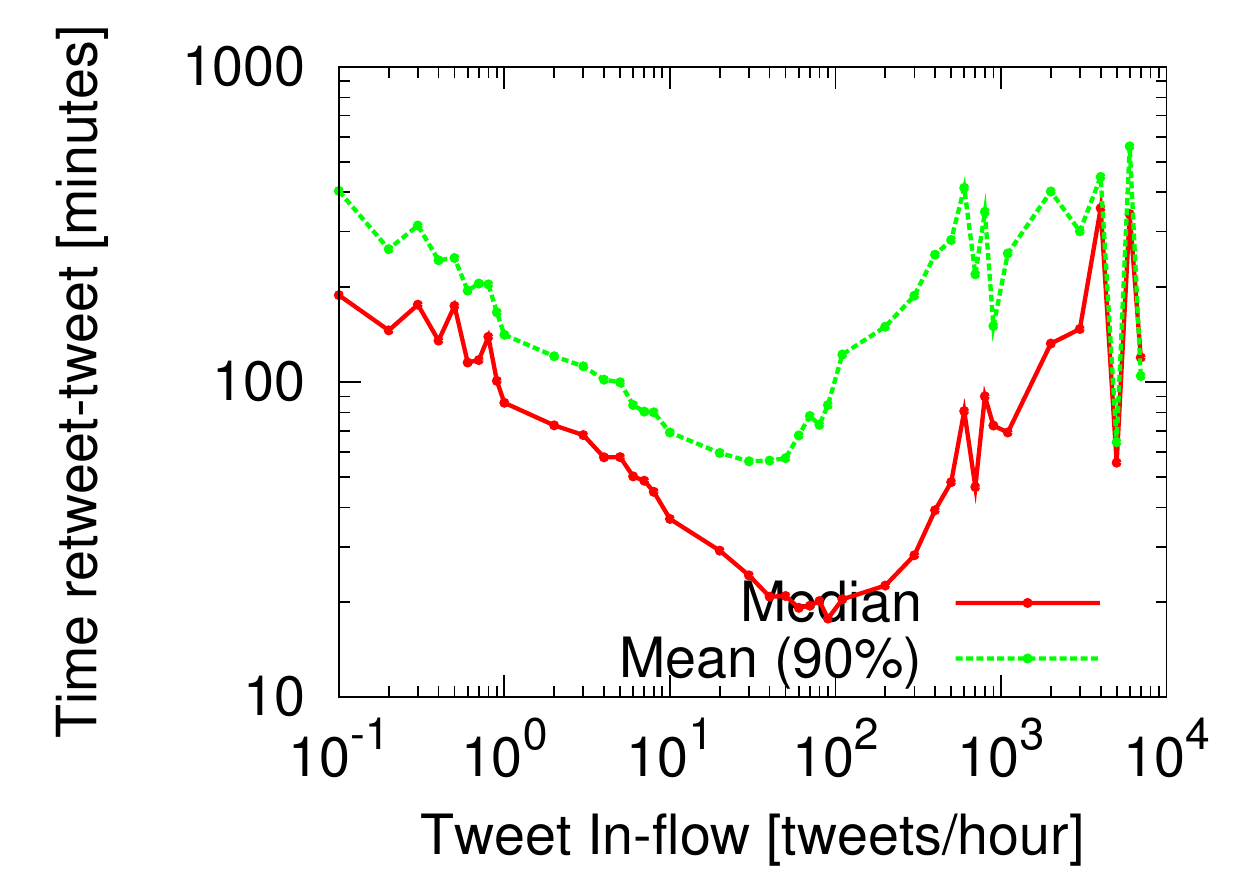}}} \hspace{2mm}
	\subfigure[Delay (non forwarded information)]{\includegraphics[width=0.28\textwidth]{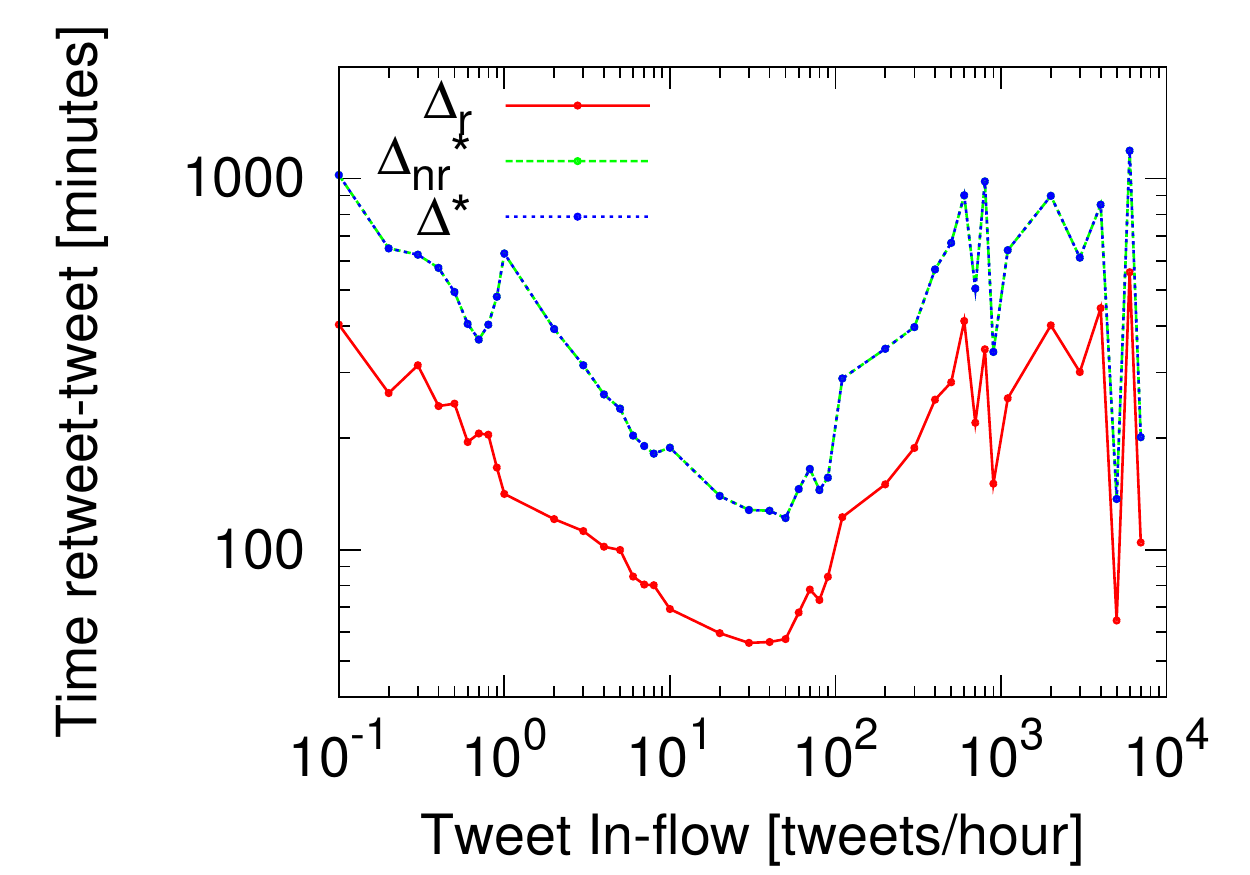} \label{fig:little-th}}
 	\caption{Queueing delay. Panel (a) shows the empirical queuing delay distribution for forwarded information. Panel (b) shows the average and median time delay between a user'{}s retweet 
	and the original tweet against in-flow rate. Panel (c) shows three average time delays for non forwarded information: $\Delta_{r}$ is the average time users take to read and decide to retweet, 
	$\Delta_{nr}^*$ is a lower bound on the average time users take to read and decide not to retweet, and $\Delta^{*}=\Delta_{r}+\Delta_{nr}^*$.} \label{fig:time-delay}
\end{figure*}

Figure~\ref{fig:queue position}(a) shows the empirical distribution of the position of a tweet on the user'{}s
queue (feed) at the time when it got retweeted for two different in-flow rate intervals, where queue position 
$q = 0$ means the tweet was at the top of the user'{}s queue at retweeting time. 
For low in-flow rates ($1-30$ tweets/hour), the user is not overwhelmed with information and is likely processing 
new tweets as soon as they arrive. That means the tweet is most likely at the head of the user'{}s queue at the time 
of retweet. 
However, for larger in-flow rates ($30-1000$ tweets/hour), the user is in\-crea\-sing\-ly overloaded and would 
likely have to process information that is accumulating in the queues faster than she can process. As a result,
we observe a shift in the most likely positions of retweets down the queue.
There is a clear difference between the plots for low and high in-flow rates in Figure~\ref{fig:queue position}(a). 
Nevertheless, in both cases, the probability of retweeting a tweet positioned beyond the first 100 slots in the queue 
is more than an order of magnitude lower than the probability of retweeting a tweet positioned in the first 10 slots in 
the queue. 
The low probability values for forwarding information located lower down the queue suggest that, for the purposes of 
information processing, queues are bounded, i.e., once information slides out of the top few positions in the queue, 
the chance of it being processed drops precipitously. 
When users are overloaded, their high tweet in-flow rates quickly push the tweets lower down the queue and beyond the 
processing limits on queue sizes.

The average (median) position of a tweet on the user'{}s queue at the time of retweet keeps increasing 
with larger in-flows, as shown in Figure~\ref{fig:queue position}(b). Interestingly, we find that the precipitous rise 
in average queue position occurs at an in-flow rate of $\sim$30 tweets/hour, matching the value of the threshold 
in-flow rate at which users begin to suffer from information overload, found previously. This finding 
suggests that when users are overloaded, their retweets are no longer drawn from their queues but instead users find 
the information they retweet through some other mechanisms. We investigate one such plausible mechanism later in
this Section.

\subsection{Quantifying queuing delays}

\noindent The temporal dynamics of information propagation, and in particular, the speed at which information propagates depends
crucially on the time users take to process the information they receive and determine if they would like to forward it to other
users~\cite{manuel11icml}.
Here, we investigate the queueing delays, i.e., the time delays between the time when a tweet was received by a user and the time when 
the user retweeted it. We are particularly interested in understanding the impact of information overload on queueing delays.

\subsubsection{Queueing delays for forwarded information}

\noindent Figure~\ref{fig:time-delay}(a) shows the empirical queueing delay distribution for di\-ffe\-rent in-flow rates. There are several
interesting patterns. First, queueing delay distributions for different in-flow rates seem to belong to the same distribution family, in
particular, the convolution of two lognormal distributions, one modeling the observation time, and another one modeling the 
reaction time, provides a good fit, in agreement with previous work~\cite{doerr2013lognormal}. 
Importantly, the mean, variance, and peak value depend on the in-flow rate. In other words, the amount of information overload of a user 
influences the time she takes to read and retweet a tweet.
The larger the tweet in-flow, the smaller time delay the peak value is located at, as shown in Fig.~\ref{fig:time-delay}(a). For exam\-ple, the
queueing delay distribution for users receiving in average $5$ to $10$ tweets/hour peaks at $5$ minutes, in contrast, the distribution for
users receiving in average $100$ to $200$ tweets per hour reaches its maximum at less $2$ minutes. This indicates that users with higher
tweet in-flow rates are more engaged into the service and observe and retweet tweets quicker.
However, although the peak value keeps shifting to smaller time values for larger in-flows, it becomes less likely and the variance increases. What about the 
median and ave\-rage time delay? Figure~\ref{fig:time-delay}(b) shows the median and (bo\-ttom 90\%) average of the empirical queueing delay against in-flow rate. 
Perhaps surprisingly, the median and average keep decreasing until some in-flow rate threshold value, and afterwards increase. Importantly, the threshold value 
for the average queueing delay seems to be coherent with our previous results, since it roughly coincides with value of the threshold in-flow rate at which users 
begin to suffer from information overload, found previously.
This suggests that overloaded users cannot keep up with the amount of in\-co\-ming information and either look for tweets directly in other 
user'{}s profiles or use tools to sort their incoming tweets.
\begin{figure*}[t]
	\centering
	\subfigure[\# sources vs in-flow rate]{\includegraphics[width=0.28\textwidth]{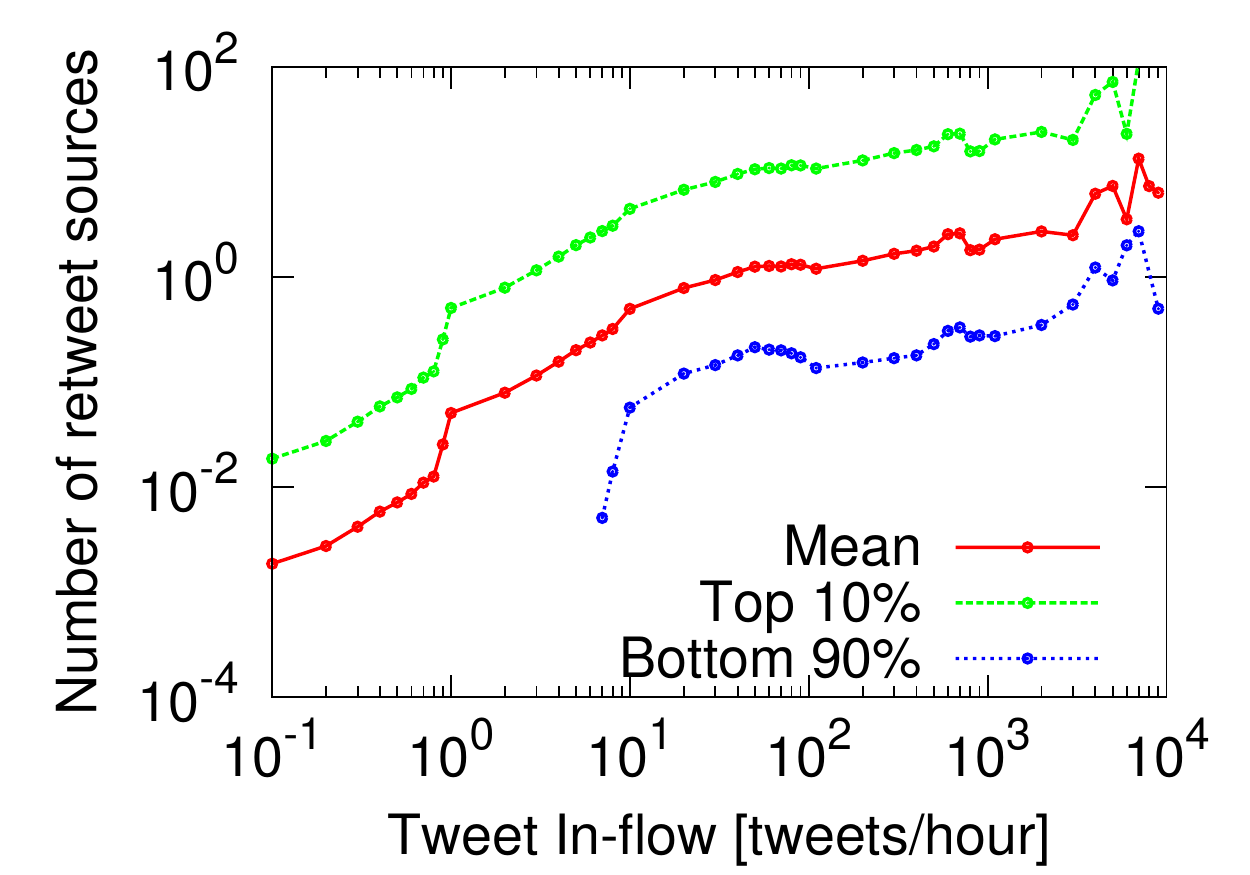}} \hspace{4mm} 
	\subfigure[\# sources vs \# followees]{\includegraphics[width=0.28\textwidth]{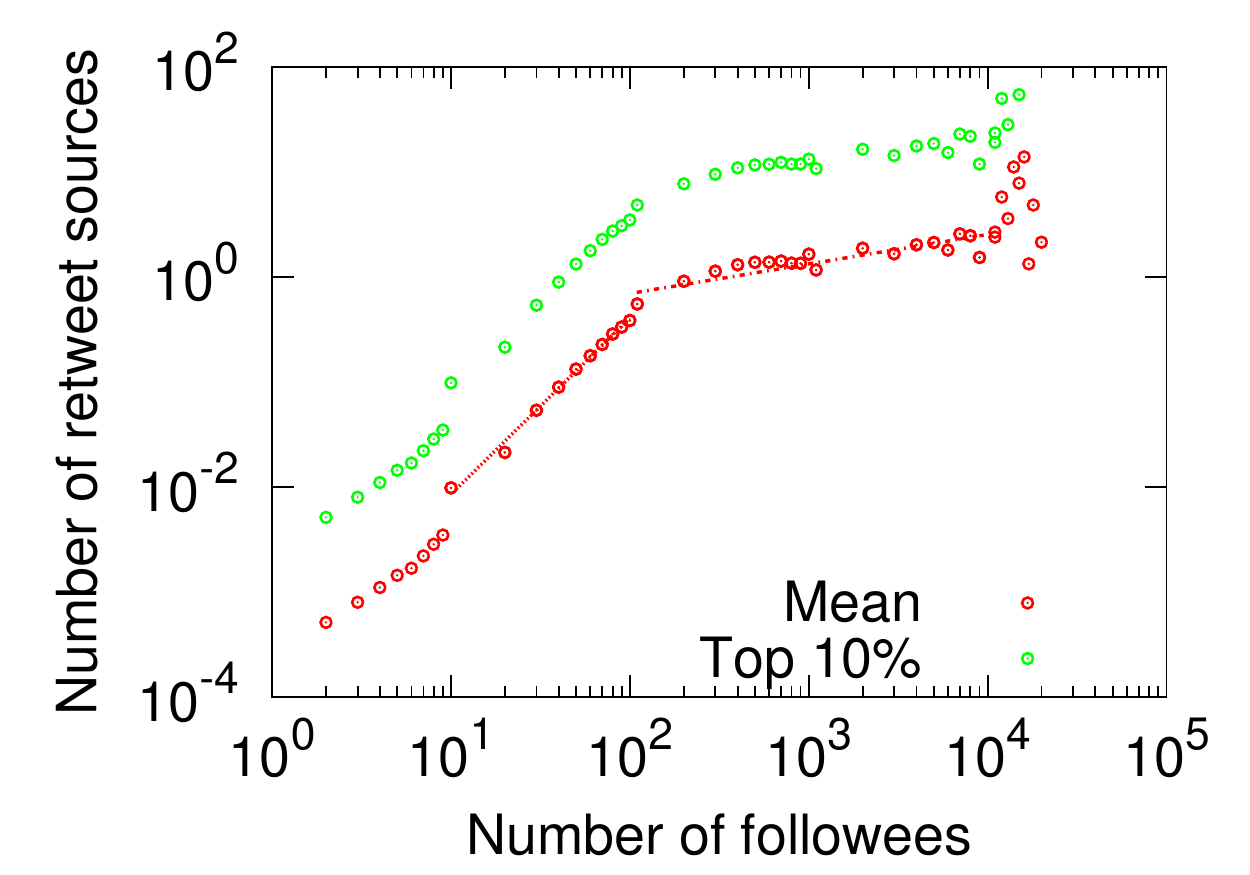}} \hspace{4mm}
	\subfigure[Source probability vs \# followees]{\includegraphics[width=0.28\textwidth]{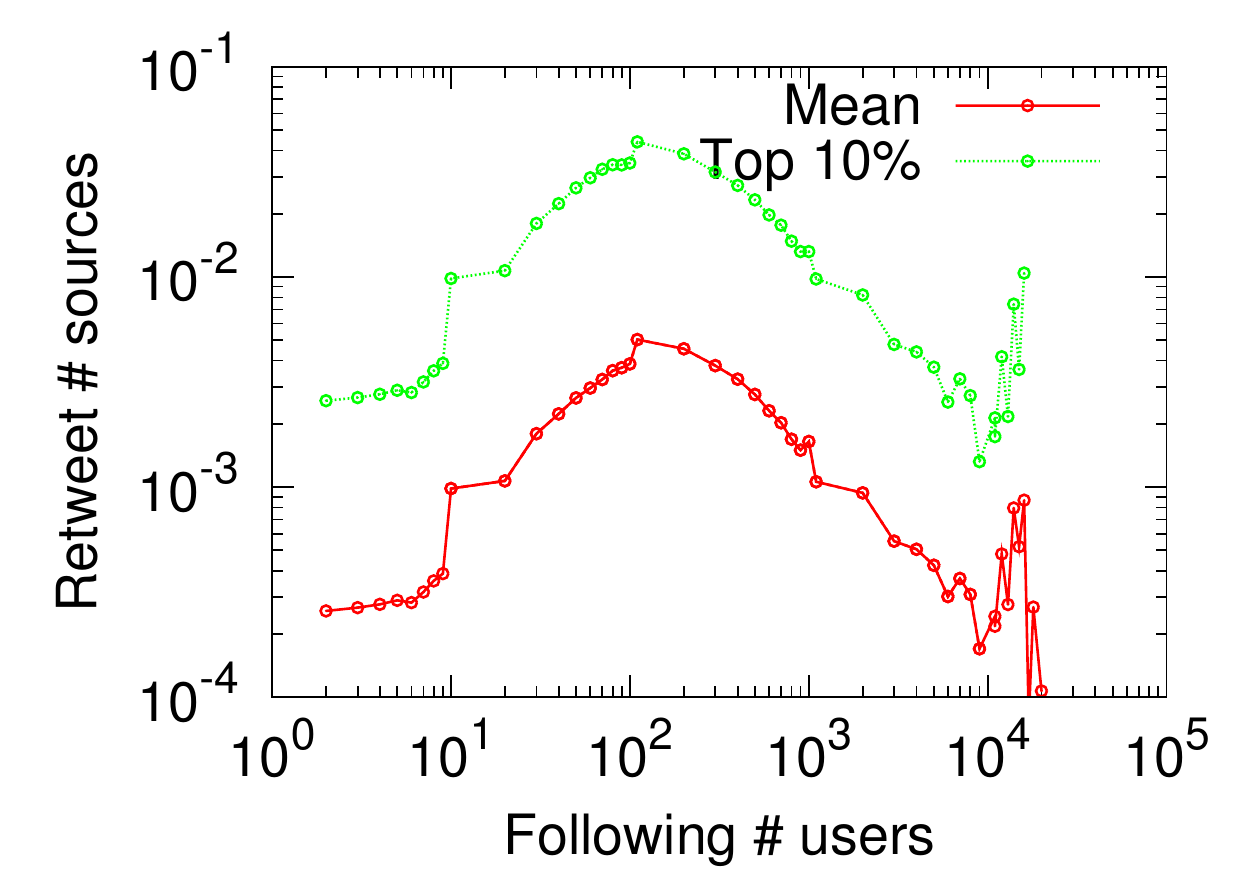}} 
 	\caption{Number of different sources vs. in-flow rate and number of followees.} \label{fig:num-sources}
\end{figure*}

\subsubsection{Queueing delays for nonforwarded information}
\noindent So far we have focused our attention on the amount of time users take to read and retweet a tweet and its position on the user'{}s queue when it got 
retweeted.
However, can we tell something about the time users take to read a tweet and decide not to retweet it? We can use Little'{}s Theorem~\cite{kleinrock1975theory} 
from queue theory to answer this question.
In our context, the theorem states that the long-term average number of unread tweets $N_{u}$ is equal to the long-term in-flow rate,
$\lambda$, multiplied by the average time a user takes to read (and possibly retweet) a tweet or average queueing delay, $\Delta$; mathematically, 
$N_{u} = \lambda \Delta$. 
Here, we assume that a non retweeted tweet \emph{exists} the queue once it exceeds a position.
We can split $\Delta$ in $(\lambda_{r} \Delta_{r} + \lambda_{nr} \Delta_{nr})/\lambda$, where $\lambda_{r}$ ($\lambda_{nr}$) 
is the in-flow rate due to tweets that (do not) get retweeted, and $\Delta_{r}$ ($\Delta_{nr}$) is the average time users take to read and decide (not) to 
retweet a tweet.
We can measure $\lambda$, $\lambda_{r}$, $\lambda_{nr}$, and $\Delta_{r}$, and bound $N_{u}$ below by the average position of a
tweet in the feed at the time it was retweeted, $N_{r} \leq N_{u}$. Therefore, we can compute a lower bound on the average time users take 
to read and decide not to retweet a tweet,
$\Delta_{nr} \geq (N_{u} - \lambda_{r} \Delta_{r})/\lambda_{nr} = \Delta^{*}_{nr}$, and thus on the average queueing delay, 
$\Delta \geq (\lambda_{r} \Delta_{r} + \lambda_{nr} \Delta^{*}_{nr})/\lambda = \Delta^{*}$.
Figure~\ref{fig:little-th} compares $\Delta_{r}$, $\Delta^{*}_{nr}$ and $\Delta^{*}$ against in-flow rate $\lambda$. Since the lower
bounds $\Delta^{*}_{nr}$ and $\Delta^{*}$ are always larger than $\Delta_{r}$, then $\Delta_{nr} \geq \Delta \geq \Delta_{r}$; in other
words, tweets that get retweeted are actually the ones that users happen to read and decide to retweet the \emph{earlier}.

It has proven difficult to know what makes an idea, a piece of information, a behavior, or, more generally, a contagion to spread quicker or slower.
Our above observations suggest that a particular social medium itself may heavily bias how quickly users process and forward information. In particular, 
it highlights the key role that a user'{}s in-flow rate in the social medium has on the user'{}s queueing delay for particular contagions on the social medium, 
independently on the contagion content. 
Later, we will investigate further the impact of a user'{}s in-flow rate on social contagion.

\subsection{Priority queueing to select sources} \label{sec:priority}
\noindent In the previous sections, we observed that when users are overloaded with information, their choice of tweets to forward becomes 
independent of the queue positions and queueing delays of the tweets. Users are likely selecting these tweets through some other mechanisms 
which differs from the LIFO information queue. 
One potential hypothesis that we investigate now is that overloaded users focus only on tweets from a small subset of all the users they follow. These 
small subset of users may be considered as influential for the overloaded users.
Then, the processing behavior of such overloaded users can be modeled using multiple prio\-ri\-ty queues~\cite{kleinrock1975theory}, where tweets from 
important sources are placed in a queue with higher priority and the rest of the tweets are placed in a lower priority queue.

To test whether overloaded users indeed limit the choice of Twitter users whose tweets they forward, we compute the 
average retweet source set size $S_{r}$\footnote{A user'{}s retweet source set is the set of different users from whom she retweets from} against in-flow 
rate and show the results in Figure~\ref{fig:num-sources}(a). We observe that si\-mi\-lar\-ly to the retweet rate, as the in-flow rate increases, the source set 
size increases at a declining rate. In other words, the source set size follows a law of diminishing returns with respect to the in-flow rate: 
$S_{r}(\lambda_1+\Delta\lambda) - S_{r}(\lambda_1) \leq S_{r}(\lambda_2+\Delta\lambda) - S_{r}(\lambda_2)$ for $\lambda_1 \geq \lambda_2$.
This sub-linear increase suggests that the larger the in-flow rate of a user, the more difficult is for a followee to become a source. 
Next, we investigate whether users tend to prioritize information from some followees over others as their number of followees increases.
Figures~\ref{fig:num-sources}(b-c) allow us to answer this question by showing the average retweet source set size and the pro\-ba\-bi\-li\-ty of a
followee to be a source against number of followees $F$. 
Interestingly, these plots exhibit two clear regimes: below $\sim$100 followees, the number of average retweet sources increases as
$S_{r} \propto F^{1.68}$, while over $\sim$100 followees, the number of retweet sources flattens dramatically, increasing slowly as
$S_{r} \propto F^{0.28}$. 
Importantly, $\sim$100 followees roughly corresponds to an in-flow rate of $\sim$30 tweets/hour, which coincides with the threshold in-flow rate at 
which users begin to suffer from information overload, found previously.
This indicates that as users follow more people and are overloaded with more information, they prioritize information produced by a smaller subset 
of influential followees, whose aggregate out-flow rate is below the users'{} processing limits.
\begin{figure*}[t]
	\centering
	\subfigure[Hashtags]{\includegraphics[width=0.28\textwidth]{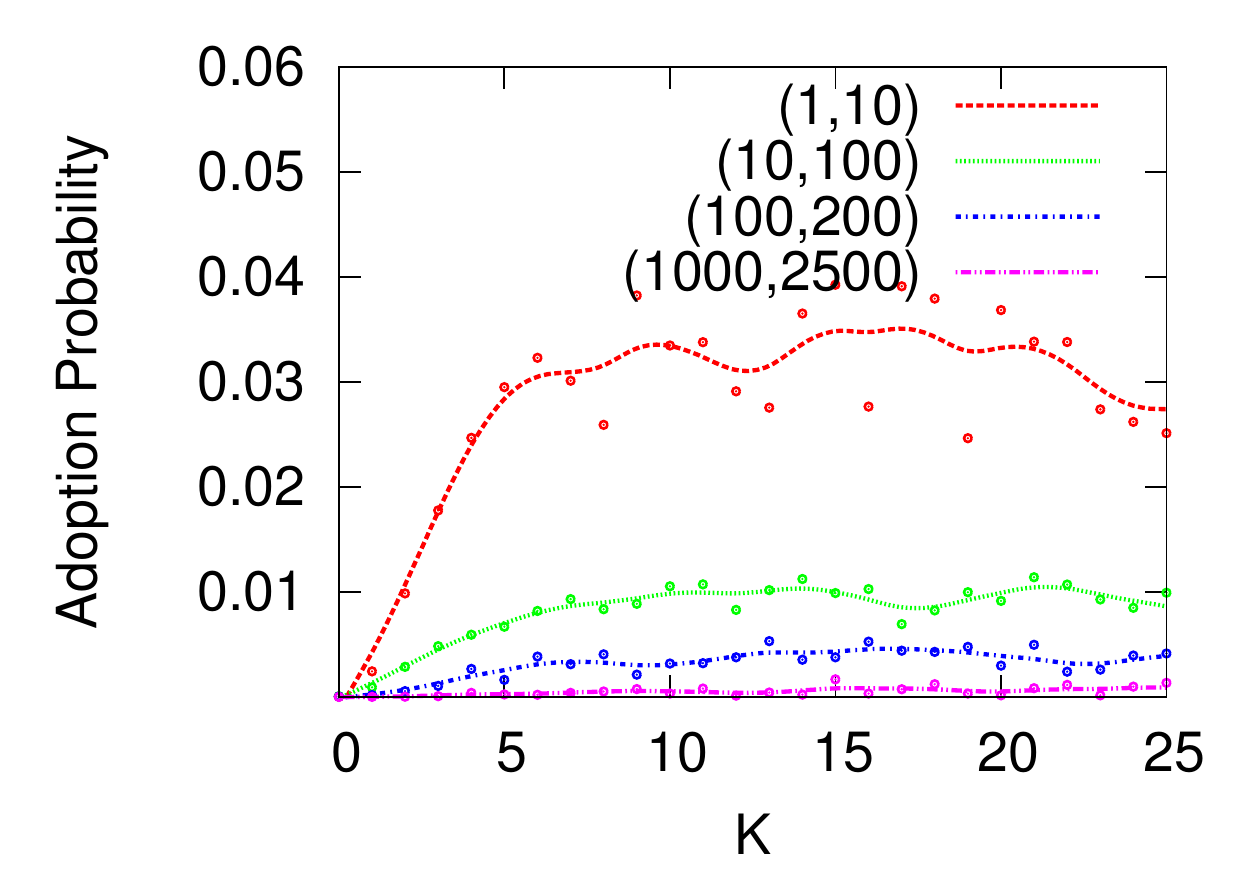} \label{fig:hashtags-in-flow}} \hspace{4mm}
 	\subfigure[``RT" retweet conventions]{\includegraphics[width=0.28\textwidth]{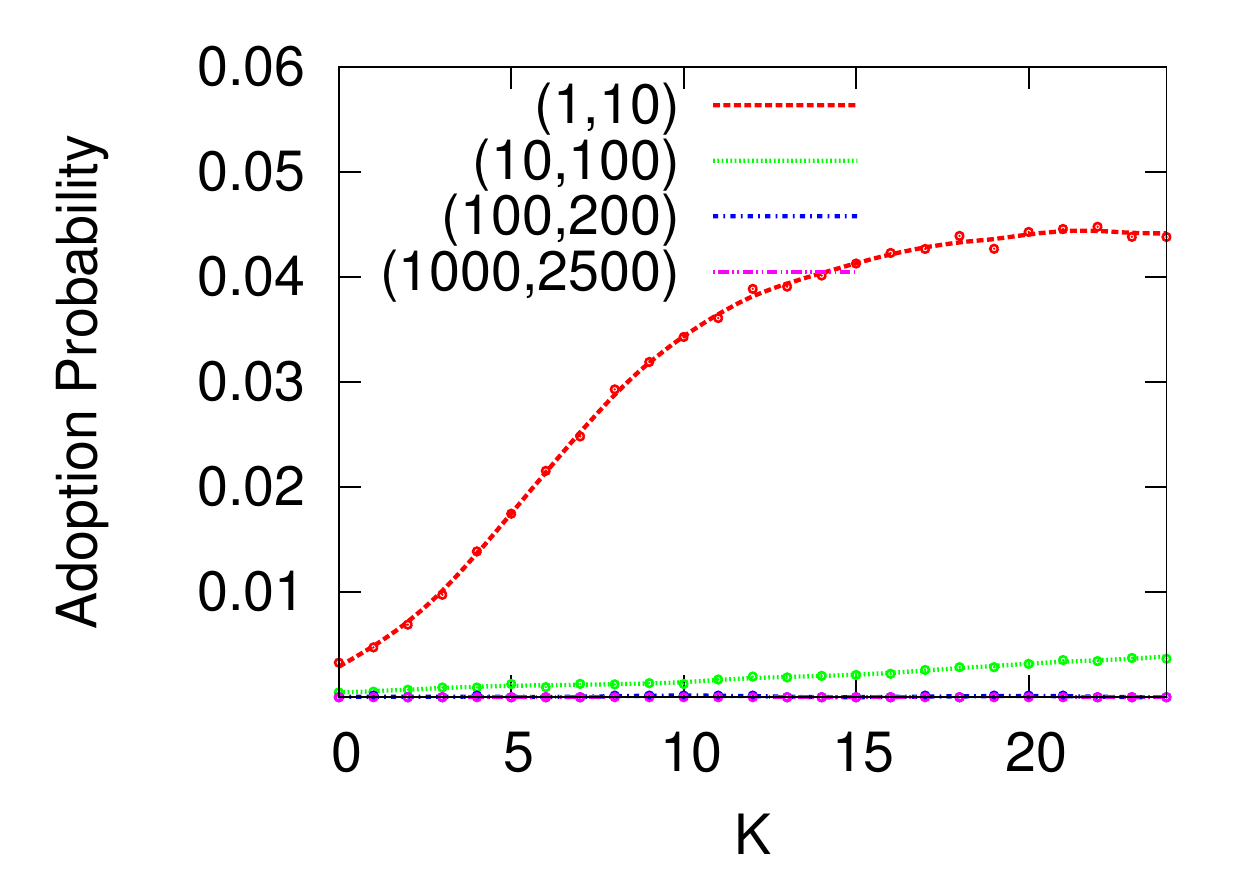} \label{fig:retweet-conventions-in-flow}} \hspace{4mm}
	\subfigure[bit.ly url shortening service]{\includegraphics[width=0.28\textwidth]{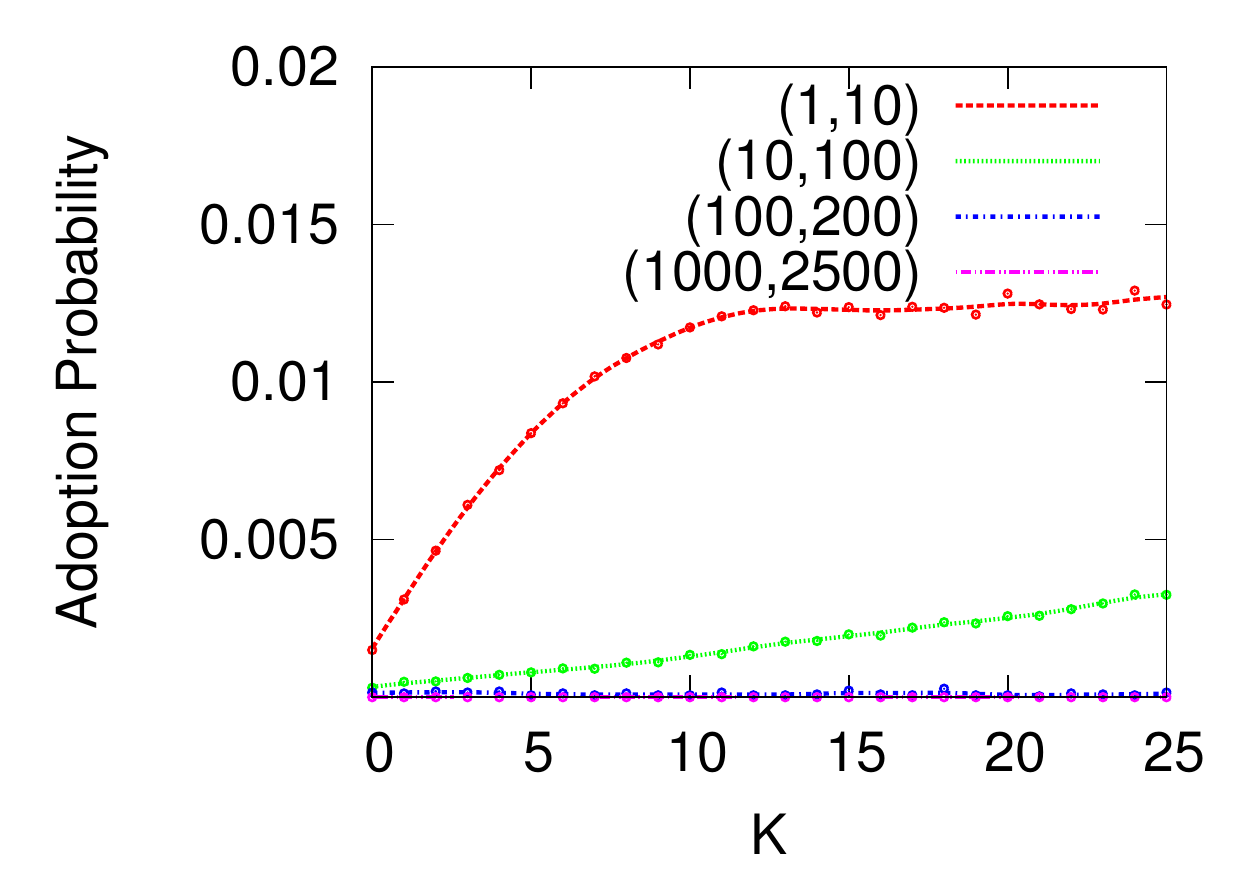} \label{fig:url-shortening-service-in-flow}}
	\caption{Exposure curves vs in-flow rate for information units (hashtags), social conventions (retweet conventions) and product adoptions (url shortening services). We group users by their 
	in-flow rate in four different ranges: $(1,10)$, $(10,100)$, $(100,200)$ and $(1000, 2500)$ tweets/hour.} \label{fig:exposure-curves-in-flow}
\end{figure*}

Our results support a previous study which analyzed mobile phone data and found that individuals exhibit a finite communication capacity,
which limits the number of ties they can maintain actively at any given time~\cite{miritello2013limited}. 
Our results also offer a different perspective on the role of influentials and the nature of their influence -- when users suffer from information
overload, they tend to prioritize or selectively process information from influential users at the expense of the remaining users. 

\section{Impact on Social Contagions}
\label{sec:adoption}
\noindent  Our study so far has demonstrated that {\it background traffic} plays an important role in users'{} decisions to forward any piece of
information they receive. Now, we investigate the impact of the background traffic and the resulting information overload on
social {\it contagions} or {\it cascades}, i.e., the viral propagation of information across a social network.

\subsection{Contagions need more exposures to spread}

\noindent In recent years, there has been an increasing interest in understanding social contagion, where a recurrent theme has been studying the number of 
exposures to a contagion a user needs to adopt it. In sociology, the
``complex contagion" principle posits that repeated exposures to an
idea are particularly crucial when the idea is in some way
controversial or
contentious~\cite{centola2010spread,centola2007complex}. Recently, 
this hypothesis has been validated in a large scale 
quantitative~\cite{romero11twitter},
and, even more recently, it has been argued that the number of social ties a user has
is crucial to assess how effective an exposure may
be~\cite{hodas2012visibility}. Here, we provide further and stronger
empirical evidence that the background traffic received by a user
plays an essential role on the number of required exposures to adopt a
contagion.

We use \emph{exposure curves}~\cite{cosley2010sequential}  
to measure the impact
that exposure to others'{} behavior has in an individual'{}s choice to
adopt a new behavior.
We say that a user is $k$-exposed to a contagion $c$ if she has not \emph{adopted} (mentioned, used) $c$, but follows $k$ other users who have 
adopted $c$ in the past. Given a user $u$ that is $k$-exposed to $c$ we would like to estimate the probability that $u$ will adopt $c$ in the future. 
Two different approaches to estimate such probability have been proposed: ordinal time estimate or snapshot estimate, the former requiring more 
detailed data~\cite{cosley2010sequential}. 
Here, we compute ordinal time estimates since our data is detailed enough, following the same procedure as~\citeauthor{romero11twitter}: 
assume that user $u$ is $k$-exposed to some contagion $c$. We estimate the probability that $u$ will adopt $c$ before becoming $(k+1)$-exposed. 
Let $E(k)$ be the number of users who were $k$-exposed to $c$ at some time, and let $I(k)$ be the number of users that were $k$-exposed and 
adopted $c$ before becoming $(k + 1)$-exposed. Then, the probability of adopting $c$ while being $k$-exposed to $c$ is 
$P(k) = I(k)/E(k)$.

We study the effects of background traffic on the exposure curves for the following three types of 
contagions over the Twitter network:

\xhdr{1. Ideas: Hashtags} Hashtags are words or phrases inside a tweet which are prefixed with the symbol \#~\cite{romero11twitter}. They provide a way for 
a user to generate searchable metadata, keywords or tags, in order to \emph{describe} her tweet, associate the tweet to a (trending) topic, or express an idea. Hashtags 
have become ubiquitous and are an integral aspect of the social Web nowadays.
Here, we consider hashtags as information units and study the impact of background traffic on their spread. In particular, we track every mention of 2,413 hashtags used 
by more than $500$ users during the three months under study. We then estimate the exposure curve of each hashtag for all users who did not use the hashtags before 
July 2009.

\xhdr{2. Conventions: Retweets} In order to study the impact of background traffic on the propagation of a social convention, we focus on the way Twitter users 
indicated back in 2009 that a tweet was being retweeted.
Different variations of this convention emerged organically during the first few years of Twitter and until November 2009, when Twitter rolled out an official, built-in retweet button. 
Here, we track every use of the most popular retweeting convention, ``RT", during the three months under study, using a similar procedure to~\citeauthor{kooti2012emergence},
and estimate exposure curves using all users who did not use ``RT" before July 2009.

\xhdr{3. Product innovations: url shortening services} We investigate the impact of background traffic on the propagation of a technological product by tracking user'{}s adoption 
of url shortening services in Twitter~\cite{antoniades2011we}.
These services existed before Twitter, however, by constraining the number of characters per message, Twitter increased their proliferation. Here, we track every 
use of the most popular url shortening service, bit.ly, during the three months under study, and estimate exposure curves using all users who did not use bit.ly before July 2009.

We analyze the influence of the background traffic on the probability of adoption of hashtags, social conventions and url shortening services by grouping users according to 
their in-flow rates and estimating an exposure curve for each group.
Figure~\ref{fig:hashtags-in-flow} shows the average exposure curve across hashtags for users with different in-flows. 
We draw se\-ve\-ral interesting observations. First, exposures for users with smaller in-flows results in a much larger increase in probability of adoption of a 
hashtag. Second, if we compare the maximum value of the probability of adoption of users with small in-flows and users with large in-flows, they differ in one 
order of magnitude.
In other words, an exposure to a hashtag is dramatically much less \emph{effective} for users that su\-ffer from information overload. 
Figure~\ref{fig:retweet-conventions-in-flow} shows exposure curves for users with different in-flows for the retweet convention ``RT". Remarkably, we find similar 
patterns to the ones we found previously for hashtags adoption Ñ an exposure to a social convention is also much less \emph{effective} for overloaded users. 
Finally, Figure~\ref{fig:url-shortening-service-in-flow} shows exposure curves for users with different in-flows for the url shortening service bit.ly. Our findings are again consistent 
with previous fin\-dings in hashtags and social conventions adoption. 

\begin{figure}[t]
	\centering
	\subfigure[Cascade size distribution]{\includegraphics[width=0.23\textwidth]{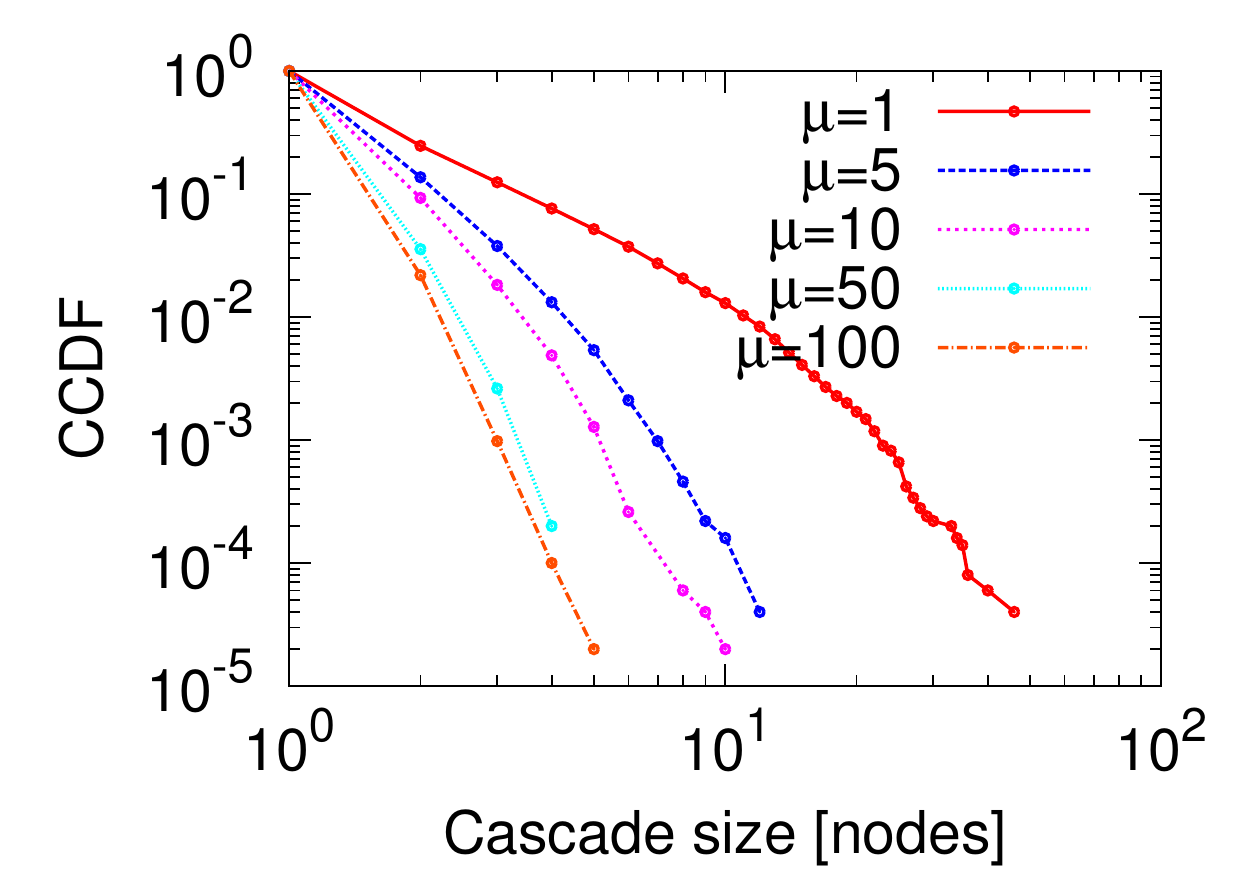}\label{fig:simulation-cascade-size}}
	\subfigure[Cascade duration distribution]{\makebox[4.2cm][c]{\includegraphics[width=0.23\textwidth]{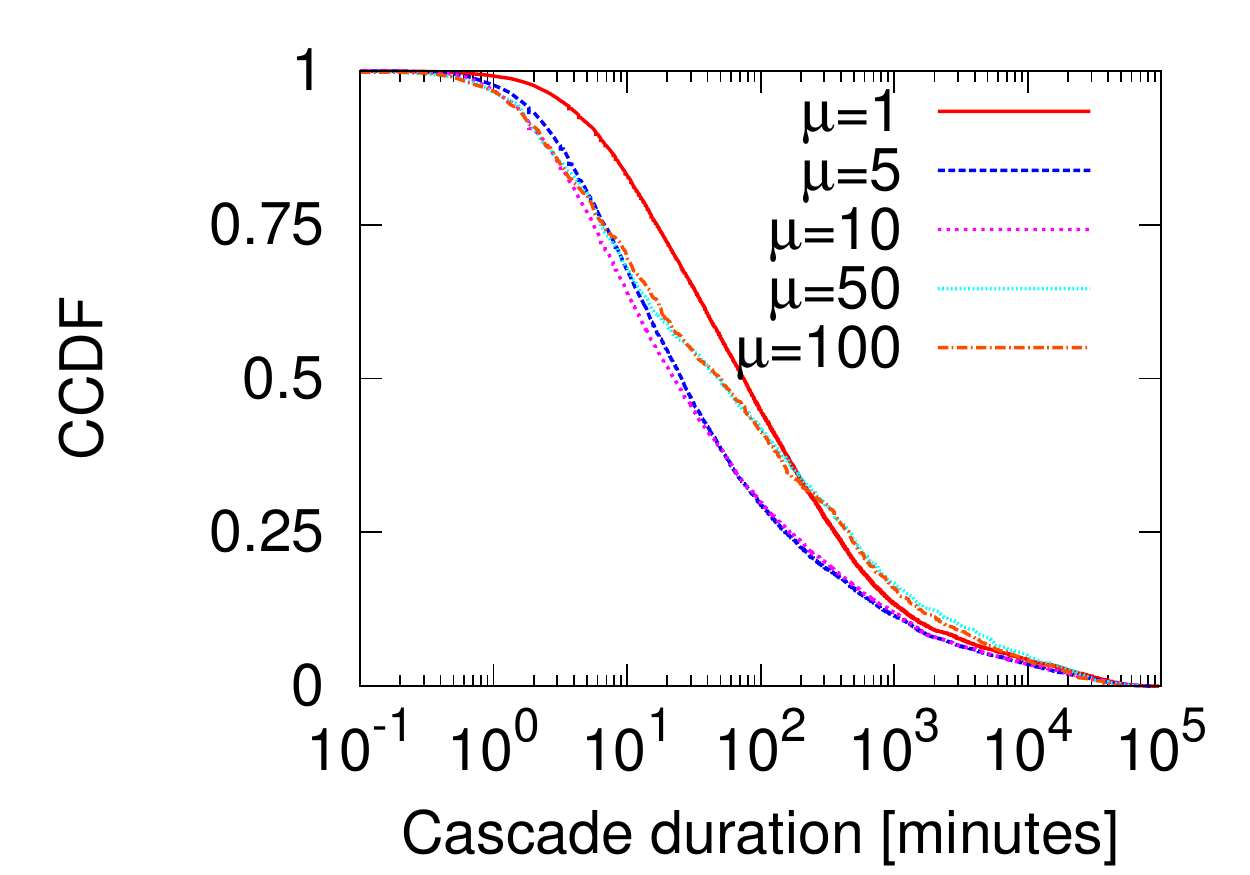}\label{fig:simulation-cascade-duration}}}
 	\caption{Cascade size and cascade duration distributions for different average in-flow rate. We used a core-periphery Kronecker network with 1,024 nodes and 20,000 edges and 
	simulated 50,000 cascades per average out-flow rate $\mu$. } \label{fig:simulation}
\end{figure}

\subsection{Cascades sizes are limited}
\noindent To the best of our knowledge, existing models of information and influence propagation do not account for the background traffic. However, we 
have given empirical evidence that the background traffic has a dramatic impact on the users'{} adoption probability in social media.
Here, we extend the well-known independent cascade model~\cite{kempe03maximizing} to support background traffic and show that, for given a network, there 
are striking differences in terms of cascade size depending on the amount of background traffic flowing through the network. The larger the background traffic, the 
shorter the cascades become.

To this aim, we first generate a network $\mathcal{G} = \left(\mathcal{V}, \mathcal{E}\right)$ using\- a well-known mathematical model of social networks: the Kronecker 
model~\cite{leskovec2010kronecker}, and set the out-flow rate of each node in the network by drawing samples from $\lambda_i \sim N(\mu, \sigma)$, where $\mu$ is the 
average out-flow rate. By this procedure, the in-flow rate for each node $j$ in the network is $\sum_{i : (i, j) \in \mathcal{E}} \lambda_i$. We then set the adoption probability 
$\beta_r $ of the independent cascade model using Fig.~\ref{fig:in-flow-probability-retweet}(b), incorporating in that way the background traffic in the model. 
Fi\-gure~\ref{fig:simulation-cascade-size} shows the cascade size distribution under different average out-flow rate values for a core-periphery Kronecker network 
(parameter matrix $[0.9, 0.5; 0.5, 0.3]$) with 1,024 nodes and 20,000 edges. We simulated 50,000 cascades per a\-ve\-rage out-flow rate.
The background traffic has a dramatic impact on the cascade size distribution, and the larger the amount of background \emph{traffic}, the more rare large cascades become. For example, 
while for $\mu = 1$, there are more than $12$\% cascades with at least $3$ nodes, for $\mu = 100$, there are only $0.1$\%.
Our simulated results provides an expla\-nation to why most information cascades in social media fail to reach epidemic proportions~\cite{ver2011stops}.

\subsection{A few cascades have prolonged lifetimes}
\noindent We have shown that the background traffic has a dramatic impact on the users'{} queueing delays, \ie, the time users take to process
information they receive. However, discrete time propagation models such us the independent cascade model used previously do not allow us to model queueing 
delays. In contrast, continuous time propagation models allow us to do so.
Here, we extend the continuous time propagation model recently introduced by~\citeauthor{manuel10netinf} to support background traffic
and show that, given a fixed network structure, there are striking differences in terms of cascade temporal duration depending on the amount of background traffic flowing 
through the network. 
In particular, we show that while most simulated cascades have shorter lifetime as the background traffic increases, as one could expect due to smaller cascade sizes, larger 
background traffic also leads to the emergence of more cascades with very prolonged lifetimes, which may be surprising at first.
However, this is a consequence of our findings in previous sections, users that suffer from information overload sometimes look for tweets directly in other user'{}s profiles 
or use tools to sort their incoming tweets and then prolong the lifetime of contagions that otherwise would die out earlier.

We proceed similarly as in previous section: we generate a synthetic network $\mathcal{G}$ using the Kronecker 
model, and set the out-flow rate of each node in the network by drawing samples from $\lambda_i \sim N(\mu, \sigma)$. 
Then, given the in-flow rate of a node $j$, we incorporate the background traffic into the continuous time model by setting the node'{}s retweet 
delay distribution $f(\Delta_r)$ using Fig.~\ref{fig:time-delay}(a) and its adoption probability $\beta_r$ using Fig.~\ref{fig:in-flow-probability-retweet}(b).
Then, we simulate and record sets of propagating cascades for different $\mu$ values.
Figure~\ref{fig:simulation-cascade-duration} shows the cascade duration distribution for cascades with $2$ or more nodes on a core-periphery Kronecker network with 1,024 nodes 
and 20,000 edges under different average out-flow rate values. We simulated $50,000$ cascades per a\-ve\-rage out-flow rate. 
Cascades that die out quicker are more frequent the larger the average out-flow rate is. However, for sufficiently large average out-flow rate ($\mu > 10$), longer cascades emerge and 
the tail of the CCDF decreases slower. This is a consequence of our previous results: people su\-ffe\-ring from large information overload cannot keep up with the amount of incoming 
information and either look for tweets directly in other user'{}s profiles or use tools to sort their incoming tweets, delaying information diffusion.

\section{Conclusions}
\label{sec:conclusions}
\noindent We have performed a large scale quantitative study of information overload by evaluating its impact on information dissemination in 
social media.
To the best of our know\-ledge, our work is the first of its kind, it reveals many in\-te\-res\-ting insights and has important implications for the large and 
growing\- number of studies on information dissemination and social contagion.
In particular, our work cha\-rac\-te\-rizes several aspects of social media users' information processing behaviors, for example, how frequently, 
from how many sources, and how quickly people forward information. It also estimates the limits of information processing of social media users
and shows that the users'{} susceptibility to social contagion depends dramatically on the rate at which they receives information, \ie, their degree 
of information overload.

Our work also opens many interesting venues for future work.
For example, we have shown that social media users'{} information
processing behavior depends on the rate at which they receive
information.
An open question is however whe\-ther it is possible to improve their
own user experience by prioritizing pieces of information from some of
their followees and particular topics over others.
Further, we have assumed information flow rates to be stationary and
the social graph to be static. However, on one hand, unexpected real
world events may trigger sudden changes in the rates of information
flows and, on the other hand, social media users'{} may start
following new users over time.
Therefore, a natural follow-up would be extending our analysis to dynamic
information flows and time-varying networks, and
investigate whether social media users modify their information
processing behavior over time.
Finally, our results rely on data gathered exclusively from
Twitter. It would be interesting to study information overload in other microblogging services (Weibo) 
and social networking sites (Facebook, G+). 

\small
\bibliographystyle{aaai}
\bibliography{refs}

\begin{thebibliography}{}

\bibitem[\protect\citeauthoryear{Antoniades \bgroup et al\mbox.\egroup
  }{2011}]{antoniades2011we}
Antoniades, D.; Polakis, I.; Kontaxis, G.; Athanasopoulos, E.; Ioannidis, S.;
  Markatos, E.~P.; and Karagiannis, T.
\newblock 2011.
\newblock we. b: The web of short urls.
\newblock {\em WWW}.

\bibitem[\protect\citeauthoryear{Backstrom \bgroup et al\mbox.\egroup
  }{2011}]{backstrom2011center}
Backstrom, L.; Bakshy, E.; Kleinberg, J.~M.; Lento, T.~M.; and Rosenn, I.
\newblock 2011.
\newblock Center of attention: How facebook users allocate attention across
  friends.
\newblock {\em ICWSM}.

\bibitem[\protect\citeauthoryear{Bawden and Robinson}{2009}]{bawden2009}
Bawden, D., and Robinson, L.
\newblock 2009.
\newblock {The dark side of information: overload, anxiety and other paradoxes
  and pathologies}.
\newblock {\em Journal of information science} 35(2):180--191.

\bibitem[\protect\citeauthoryear{Bontcheva, Gorrell, and
  Wessels}{2013}]{bontcheva2013}
Bontcheva, K.; Gorrell, G.; and Wessels, B.
\newblock 2013.
\newblock {Social Media and Information Overload: Survey Results}.
\newblock {\em Arxiv preprint arXiv:1306.0813}.

\bibitem[\protect\citeauthoryear{Borchers \bgroup et al\mbox.\egroup
  }{1998}]{borchers1998}
Borchers, A.; Herlocker, J.; Konstan, J.; and Reidl, J.
\newblock 1998.
\newblock {Ganging up on information overload}.
\newblock {\em Computer} 31.

\bibitem[\protect\citeauthoryear{Centola and Macy}{2007}]{centola2007complex}
Centola, D., and Macy, M.
\newblock 2007.
\newblock Complex contagions and the weakness of long ties.
\newblock {\em American Journal of Sociology} 113(3):702--734.

\bibitem[\protect\citeauthoryear{Centola}{2010}]{centola2010spread}
Centola, D.
\newblock 2010.
\newblock The spread of behavior in an online social network experiment.
\newblock {\em Science} 329(5996):1194--1197.

\bibitem[\protect\citeauthoryear{Cha \bgroup et al\mbox.\egroup
  }{2010}]{cha2010measuring}
Cha, M.; Haddadi, H.; Benevenuto, F.; and Gummadi, P.~K.
\newblock 2010.
\newblock {Measuring User Influence in Twitter: The Million Follower Fallacy}.
\newblock {\em ICWSM}.

\bibitem[\protect\citeauthoryear{Comarela \bgroup et al\mbox.\egroup
  }{2012}]{comarela2012understanding}
Comarela, G.; Crovella, M.; Almeida, V.; and Benevenuto, F.
\newblock 2012.
\newblock Understanding factors that affect response rates in twitter.
\newblock {\em HT}.

\bibitem[\protect\citeauthoryear{Cosley \bgroup et al\mbox.\egroup
  }{2010}]{cosley2010sequential}
Cosley, D.; Huttenlocher, D.~P.; Kleinberg, J.~M.; Lan, X.; and Suri, S.
\newblock 2010.
\newblock Sequential influence models in social networks.
\newblock {\em ICWSM}.

\bibitem[\protect\citeauthoryear{Dean and Webb}{2011}]{mckinsey2011}
Dean, D., and Webb, C.
\newblock 2011.
\newblock {Recovering from information overload}.
\newblock {\em McKinsey Quarterly}.

\bibitem[\protect\citeauthoryear{Doerr, Blenn, and
  Van~Mieghem}{2013}]{doerr2013lognormal}
Doerr, C.; Blenn, N.; and Van~Mieghem, P.
\newblock 2013.
\newblock Lognormal infection times of online information spread.
\newblock {\em PloS one} 8(5):e64349.

\bibitem[\protect\citeauthoryear{Du \bgroup et al\mbox.\egroup
  }{2013}]{du13nips}
Du, N.; Song, L.; Gomez-Rodriguez, M.; and Zha, H.
\newblock 2013.
\newblock Scalable influence estimation in continuous-time diffusion networks.
\newblock {\em NIPS}.

\bibitem[\protect\citeauthoryear{Goel, Watts, and
  Goldstein}{2012}]{goel2012structure}
Goel, S.; Watts, D.~J.; and Goldstein, D.~G.
\newblock 2012.
\newblock The structure of online diffusion networks.
\newblock {\em {EC}}.

\bibitem[\protect\citeauthoryear{Gomez-Rodriguez, Balduzzi, and
  Sch\"{o}lkopf}{2011}]{manuel11icml}
Gomez-Rodriguez, M.; Balduzzi, D.; and Sch\"{o}lkopf, B.
\newblock 2011.
\newblock {Uncovering the Temporal Dynamics of Diffusion Networks}.
\newblock {\em ICML}.

\bibitem[\protect\citeauthoryear{Gomez-Rodriguez, Leskovec, and
  Krause}{2010}]{manuel10netinf}
Gomez-Rodriguez, M.; Leskovec, J.; and Krause, A.
\newblock 2010.
\newblock {Inferring Networks of Diffusion and Influence}.
\newblock {\em KDD}.

\bibitem[\protect\citeauthoryear{Goyal and Kearns}{2012}]{goyal2012competitive}
Goyal, S., and Kearns, M.
\newblock 2012.
\newblock Competitive contagion in networks.
\newblock {\em STOC}.

\bibitem[\protect\citeauthoryear{Gross}{1964}]{gross1964managing}
Gross, B.~M.
\newblock 1964.
\newblock {\em The managing of organizations: The administrative struggle}.
\newblock Free Press of Glencoe New York.

\bibitem[\protect\citeauthoryear{Hodas and Lerman}{2012}]{hodas2012visibility}
Hodas, N., and Lerman, K.
\newblock 2012.
\newblock How visibility and divided attention constrain social contagion.
\newblock {\em SocialCom}.

\bibitem[\protect\citeauthoryear{Hodas, Kooti, and Lerman}{2013}]{hodas13icwsm}
Hodas, N.; Kooti, F.; and Lerman, K.
\newblock 2013.
\newblock Friendship paradox redux: your friends are more interesting than you.
\newblock {\em ICWSM}.

\bibitem[\protect\citeauthoryear{Kempe, Kleinberg, and
  Tardos}{2003}]{kempe03maximizing}
Kempe, D.; Kleinberg, J.~M.; and Tardos, E.
\newblock 2003.
\newblock Maximizing the spread of influence through a social network.
\newblock {\em KDD}.

\bibitem[\protect\citeauthoryear{Kleinrock}{1975}]{kleinrock1975theory}
Kleinrock, L.
\newblock 1975.
\newblock {\em Theory, volume 1, Queueing systems}.
\newblock Wiley-interscience.

\bibitem[\protect\citeauthoryear{Kooti \bgroup et al\mbox.\egroup
  }{2012}]{kooti2012emergence}
Kooti, F.; Yang, H.; Cha, M.; Gummadi, P.~K.; and Mason, W.~A.
\newblock 2012.
\newblock The emergence of conventions in online social networks.
\newblock {\em ICWSM}.

\bibitem[\protect\citeauthoryear{Kwak \bgroup et al\mbox.\egroup
  }{2010}]{kwak2010twitter}
Kwak, H.; Lee, C.; Park, H.; and Moon, S.
\newblock 2010.
\newblock What is twitter, a social network or a news media?
\newblock {\em WWW}.

\bibitem[\protect\citeauthoryear{Leskovec, Backstrom, and
  Kleinberg}{2009}]{leskovec2009kdd}
Leskovec, J.; Backstrom, L.; and Kleinberg, J.
\newblock 2009.
\newblock Meme-tracking and the dynamics of the news cycle.
\newblock {\em KDD}.

\bibitem[\protect\citeauthoryear{Leskovec \bgroup et al\mbox.\egroup
  }{2010}]{leskovec2010kronecker}
Leskovec, J.; Chakrabarti, D.; Kleinberg, J.; Faloutsos, C.; and Ghahramani, Z.
\newblock 2010.
\newblock {Kronecker graphs: An approach to modeling networks}.
\newblock {\em JMLR}.

\bibitem[\protect\citeauthoryear{Miritello \bgroup et al\mbox.\egroup
  }{2013}]{miritello2013limited}
Miritello, G.; Lara, R.; Cebrian, M.; and Moro, E.
\newblock 2013.
\newblock Limited communication capacity unveils strategies for human
  interaction.
\newblock {\em Scientific reports} 3.

\bibitem[\protect\citeauthoryear{Mislove \bgroup et al\mbox.\egroup
  }{2007}]{mislove2007measurement}
Mislove, A.; Marcon, M.; Gummadi, K.~P.; Druschel, P.; and Bhattacharjee, B.
\newblock 2007.
\newblock Measurement and analysis of online social networks.
\newblock {\em {SIGCOMM}}.

\bibitem[\protect\citeauthoryear{Myers and Leskovec}{2012}]{myers12clash}
Myers, S., and Leskovec, J.
\newblock 2012.
\newblock Clash of the contagions: Cooperation and competition in information
  diffusion.

\bibitem[\protect\citeauthoryear{Romero, Meeder, and
  Kleinberg}{2011}]{romero11twitter}
Romero, D.~M.; Meeder, B.; and Kleinberg, J.
\newblock 2011.
\newblock Differences in the mechanics of information diffusion across topics:
  idioms, political hashtags, and complex contagion on twitter.
\newblock {\em WWW}.

\bibitem[\protect\citeauthoryear{Toffler}{1984}]{toffler1984future}
Toffler, A.
\newblock 1984.
\newblock {\em Future shock}.
\newblock Random House.

\bibitem[\protect\citeauthoryear{Ver~Steeg, Ghosh, and
  Lerman}{2011}]{ver2011stops}
Ver~Steeg, G.; Ghosh, R.; and Lerman, K.
\newblock 2011.
\newblock What stops social epidemics?
\newblock {\em ICWSM}.

\bibitem[\protect\citeauthoryear{Weng \bgroup et al\mbox.\egroup
  }{2012}]{weng2012competition}
Weng, L.; Flammini, A.; Vespignani, A.; and Menczer, F.
\newblock 2012.
\newblock Competition among memes in a world with limited attention.
\newblock {\em Scientific Reports} 2.

\end{thebibliography}

\end{document}